\documentclass[11pt]{article}
\usepackage{graphicx}
\usepackage{booktabs}
\usepackage{array, amsmath, graphicx, mathtools, amssymb, mathtools, setspace, booktabs, float, lscape, xcolor, authblk}
\usepackage[title]{appendix}
\usepackage{ caption, txfonts, pdflscape, fixltx2e, float}

\usepackage[labelfont = sf]{subcaption}

\usepackage[authoryear, round]{natbib}
\graphicspath{{FIGURES/}}

\usepackage{url, makeidx}

\usepackage{ wrapfig, subcaption, tikz, enumitem} 

\newcommand{\W}{\hphantom{0}}

\usepackage{xspace}

\begin{document}
\title{Modulation and integration in a local processor having input from two distinct sources}

\author[1]{Jim W. Kay}
\author[2]{Bruce P. Graham}
\affil[1]{School of Mathematics and Statistics, University of Glasgow; jim.kay@glasgow.ac.uk}
\affil[2] {Computing Science and Mathematics, Faculty of Natural Sciences, University of Stirling; bruce.graham@stirling.ac.uk}

\maketitle

\begin{abstract}
By making use of partial information decomposition, we analyse information transmission through a local processor which receives two streams of integrated inputs. We revise previously stated conditions for contextual modulation that are based on the components of partial information decomposition, and introduce a condition for the presence of signal integration. The conditional probability for the occurrence of an event, such as a spike or burst, is defined based on a general activation function of two distinct integrated inputs to the local processor. Exact results are obtained for information-theoretic measures and the components of partial information decomposition using the Ibroja method. These results are applied with several activation functions, of additive, multiplicative, subtractive and divisive forms with graphical illustration. We find that transfer functions based on multiplicative activation provide modulatory behaviour, but when the activation function is a simple product of the inputs this is not the case, and the behaviour is integrative. When the activation function is a simple addition of the two inputs we find by considering restricted domains that modulatory behaviour of two types is apparent. Furthermore, when both inputs are large and increasing then signal integration occurs in the form of increasing synergy and shared information. Similar signal integration occurs when the activation function is a simple subtraction of the two inputs, and modulatory behaviour occurs in one of the restricted domains but not the other. A form of divisive normalisation transfer function is also considered and it turns out to be purely modulatory, with integrated input from neighbouring cells providing the modulation of the driving input. The revised information-theoretic conditions are also applied to noisy data obtained from a realistic simulation of a model layer 5b pyramidal cell in three distinct regimes defined by experimental design and characterised by the values of the input basal and apical amplitudes. Evidence for driving basal input with modulating apical input, driving apical input with modulating basal input, and of apical-basal signal integration is presented.
\end{abstract}

\W\W\W\W\W\W\W\W\W\W  transfer function,  layer 5b pyramidal cell, apical amplification, apical drive,    \\  \W\W\W\W\W\W\W\W\W\W  synergy, subtractive inhibition, divisive normalisation
\section*{Introduction} 
We consider a local processor which receives input from two distinct sites of integration, namely (a) a receptive field (RF) and (b) a contextual field (CF). Such a processor can be viewed as being part of an artificial neural network \citep{PKS}  or as being a layer 5b pyramidal cell in which the sites of integration lie in the basal and apical dendrites, respectively \citep{Lark}.
In earlier work \citep{KIDP, KP}, partial information decomposition (PID) \citep{WB} was used to separate out different components of the transmitted information. These components were  employed in the definition of contextual modulation. The focus of the earlier work was decidedly asymmetric in that the integrated receptive field input was assumed to be driving and the only role for the contextual input was to amplify or attenuate the transmission of information.  The possibility that the integrated contextual field input could be driving, and even be modulated by the integrated receptive field input, was ruled out. This stipulation was enforced by a specially-crafted modulatory activation function \citep{PKS}. Use of partial information decomposition demonstrated that this and other  modulatory activation functions enable  particular forms of information processing that are very different from those obtained by employing  activation functions that involve simple addition, subtraction, multiplication or divisive combinations of the two inputs to the processor \citep{KP}. For reviews of the occurrence of contextual modulation in neuroscience, see \citet{RAS}, \citet{PCS} and \citet{WAP}. 

Recent work \citep{GKP}, in which a reduced, but realistic, model of a layer 5b pyramidal cell was used to generate data regarding the occurrence of  a burst, given basal and apical step currents of varying amplitudes, has included the possibility that a burst could be generated by (a) basal input alone, (b) apical input alone (c) a combination of basal and apical input -- or not at all.  These four distinct regimes were implemented by careful experimental design. 

In the sequel, we speak of drive and context rather than RF and CF and take a more neutral approach in which either of the two inputs, or both, can be driving and the other modulating. When we speak of ``inputs'' in the sequel we actually mean ``integrated'' inputs. Given two distinct binary inputs and a binary output, we define a class of transfer functions, in which each function maps  the strengths of the two inputs  to the probability of an event at the output. For a given transfer function and a fixed combination of signal strengths, a trivariate binary probability distribution is defined. Partial information decomposition is then applied to this probability distribution. This is repeated for many combinations of the signal strengths of the two inputs, and the PID components provide the pattern of information transmission in each case. We provide exact results for the Ibroja PID \citep{GK, Bert}  in which the PID components are obtained as continuous functions of the signal strengths of the two inputs. We also define a transfer function which implements a form of divisive normalisation \citep{CH},  and we derive exact results for the Ibroja PID components also for this form of transfer function.
We also revise the previous conditions for contextual modulation and define a condition for signal integration. By performing tests based on comparing various combinations of input signal strengths, these conditions are checked by employing a number of transfer functions of modulatory and integrative types. 

We also consider data on the occurrence of a burst for given levels of basal and apical amplitude using simulations of  a model pyramidal layer 5b cell \citep{GKP} and 
 reframe the earlier work on contextual modulation in terms of integrated fields of basal and apical inputs, respectively.  These conditions for contextual modulation are then checked  for the regimes where basal input amplitude is high while apical input amplitude is low, and vice versa, by using a numerical algorithm to obtain the Ibroja PID \citep{JEC}  for combinations of input basal and apical amplitudes.  For the regime in which both basal and apical input amplitudes are high, we demonstrate the integration of the basal and apical inputs in producing a burst.

\section*{Methods}

\subsection*{Notation}
We consider the trivariate probability distribution involving binary random variables, $Y, X_1,$
$ X_2$, with probability mass function (p.m.f.)  $p(y, x_1, x_2)$. The variable $Y$  is an output variable with levels $0$ and $1,$ where level 1 denotes the occurrence of an event, e.g. a spike or a burst within a time interval of given length.  The  input, $X_1$, has two levels: $0$ and $s_1$, where $s_1$ is strictly positive, and it denotes the strength of signal $X_1$ when present. The  input, $X_2$, has two levels: $0$ and $s_2$, where $s_2$ is strictly positive, and it denotes the strength of signal $X_2$ when present. We will make use of standard information measures, such as entropy, mutual information and conditional mutual information etc. \citep{CT}.  See \citet{WPKLP} for a description of these measures and also partial information decomposition.

\subsection*{Partial information decomposition (PID)}

We can write the Shannon entropy $H(Y)$ as
\begin{equation*}
H(Y) = I(Y; X_1, X_2) + H(Y)_\text{res},
\end{equation*}
where $H(Y)_\text{res}$ is the conditional output entropy $H(Y | X_1, X_2)$. Then the joint mutual information (JMI) $I(Y; X_1,X_2)$ may be decomposed into four non-negative information components  \citep{WB} as. 
\begin{equation*}
I(Y ; X_1, X_2) = \text{Unq}X_1 + \text{Unq}X_2 + \text{Shd} + \text{Syn},
\end{equation*}
where these terms describe parts of the information transmitted by the output $Y$. They are the unique information Unq$X_1$ that input, $X_1,$
possesses regarding the output, that is not available from input $X_2$; the unique information Unq$X_2$ that input, $X_2,$
possesses regarding the output, that is not available from input $X_1;$ the shared information Shd that both inputs possess regarding the output and  the synergistic information Syn that the inputs in combination have about the output. In the following we will use the Ibroja PID \citep{GK, Bert}. Typically, application of this PID involves numerical optimisation, but for the class of transfer functions considered here we show that there are exact formulae for each of the PID components and that they can be written as functions of the two signal strengths $(s_1, s_2).$ Since we will be comparing many different 2 by 2 by 2 PIDs, defined by  combinations of the signal strengths, we normalise the components in each PID by dividing by the output entropy, $H(Y)$. The value of $H(Y)$ gives the total bandwith -- the total amount of information that can be transmitted -- while the value of the JMI is the amount of information that is actually transmitted. In particular scenarios the transmitted information can use up all of the available bandwidth, while in others only a small proportion of the available bandwidth is utilised.

The PID component, Shd, can be split into two non-negative parts,  Shd = ShdS + ShdM, where ShdS and ShdM  are called the source shared and mechanistic shared information, respectively \citep{HSP, PPCP}. It is not possible to provide exact formulae for these components but  \citet{PPCP} provide expressions from which they can be determined numerically. In any application involving independent inputs, with $I(X_1; X_2)$ =0,  we can deduce that ShdS =0 and so ShdM = Shd \citep[Eq. (9)]{PPCP}, so that the exact Shd component provides the exact mechanistic shared information transmitted by the system.

\subsection*{Conditions for contextual modulation}
The conditions for contextual modulation are based on \citet{KP}, with modification. 
They are chosen to test for the key properties of contextual modulation.  In applications, one of the two inputs will provide {\em the drive} and the other {\em the context}. The conditions are: 
\begin{enumerate}
\item[CM1:] The drive is sufficient for the transmission of the information that the inputs possess about the output, so the context is not necessary.
\item[CM2:] The drive is necessary for the  transmission of the information that the inputs possess about the output, so the context is not sufficient.
\item[CM3:] The output can transmit unique information provided by the drive but little or no unique information  provided by the context. 
\item[CM4:] When the drive is weak, the context can strengthen the transmission of the summed information components: unique information provided by the drive, shared information and synergistic information.
\end{enumerate}

Four combinations of signal strengths are used to check these conditions: 
\begin{center}
\begin{tabular}{ll} \toprule
 S\textsubscript{1} &Scenario 1: Strong drive, with context near to 0, or absent\\
S\textsubscript{2} &Scenario 2: Drive near to zero, or absent, with  strong context \\
& \hspace{2cm} but not strong enough to generate an event by itself \\
S\textsubscript{3} &Scenario 3: Weak drive, with context near to 0, or absent \\
S\textsubscript{4} &Scenario 4: Weak drive, with moderate context \\
\bottomrule
\end{tabular}
\end{center}

The conditions CM1 and CM2 assume implicitly that the possible values for the drive and the context can be varied independently. This is certainly the case when the set of $(s_1, s_2)$ values of interest is a rectangular region. When the set of $(s_1, s_2)$ values is triangular, e.g. $s_1  > s_2$, then these variables can't be varied independently, and so conditions CM1 and CM2 cannot both be checked. If $s_1 > s_2$ only CM1 can be checked, whereas if $s_1 < s_2$ only CM2 can be checked -- similarly for the corresponding scenarios S\textsubscript{1} and S\textsubscript{2}. 

Overall, we state that contextual modulation occurs because the context can contribute to the transmission of information via the unique component provided by the drive along with the shared and synergistic components when the drive is weak, while little or no unique information provided by the context  itself is transmitted.

\subsection*{Condition for the integration of inputs}
We define a condition for integration of the inputs in the case where both inputs are high: {\em when both inputs are sufficiently high then most information is transmitted via the shared and synergistic components,  with little or no unique information  provided by either input being transmitted.}

\subsection*{Probability specifications}
The random variables $X_1, X_2$ are assumed to each follow the uniform binary distribution. The marginal distribution of $(X_1, X_2)$ is given in Table~\ref{tabRC}.
\begin{table}[H]
\centering
\caption{Marginal probability distribution of the  inputs, $X_1, X_2$} \label{tabRC}
\begin{tabular}{ccc} \toprule
$x_1$ & $x_2$ & $p(x_1, x_2)$ \\ \midrule
0 & 0 &$ \lambda = (1 + \rho)/4$ \\
$s_1$ & 0  & $\mu = (1 - \rho)/4$ \\
0& $s_2$ & $\mu = (1 - \rho)/4$ \\
$s_1$ & $s_2$ & $\lambda = (1 + \rho)/4$ \\
\bottomrule
\end{tabular}
\end{table}
\noindent where $\rho$ is the correlation between the inputs $X_1, X_2 \, (-1 < \rho <1)$ and $s_1 >0, s_2>0$.  We write the p.m.f. for $(Y, X_1, X_2)$ as
\begin{equation}
p(y, x_1, x_2)= \Pr(Y= y | X_1 = x_1, X_2 = x_2) p(x_1, x_2) \label{pyrc}
\end{equation}
We specify the conditional output probabilities that $Y=1$, given the inputs, $X_1, X_2$,  in Table~\ref{tabCOP}. Let $F: \mathbb{R}\mapsto (0,1)$ be a strictly increasing, continuous cumulative probability distribution function, where $\mathbb{R}$ is the set of real numbers. Let the function $a:\mathbb{R}_{\geq 0} \times \mathbb{R}_{\geq 0}\mapsto \mathbb{R}_{\geq 0}$ denote the combined activation of the two inputs, $X_1, X_2$, where $\mathbb{R}_{\geq 0}$ denotes the set of non-negative real numbers. Then the transfer function is defined to be
\begin{equation}
T(x_1, x_2) = F( a(x_1, x_2) - \tau), \label{TF}
\end{equation}
where $\tau$ is a fixed threshold parameter. For simplicity of notation, we make the following identifications:
\begin{equation}
a_{00} \equiv a(0, 0), \quad a_{10} \equiv a(s_1, 0), \quad a_{02} \equiv a(0, s_2), \quad a_{12} \equiv a(s_1, s_2). \label{ident}
\end{equation}
\begin{table}[H]
\centering
\caption{Conditional output probabilities that $Y =1$ given $(X_1, X_2)$}  \label{tabCOP}
\begin{tabular}{ccc} \toprule
$x_1$ & $x_2$ & $\Pr(Y=1 | X_1 = x_1, X_2 = x_2) $ \\ \midrule
0 & 0 &$F(a_{00}-\tau)$ \\
$s_1$ & 0  & $F(a_{10}- \tau)$ \\
0& $s_2$ & $F(a_{02} -\tau)$ \\
$s_1$ & $s_2$ & $F(a_{12} - \tau)$ \\
\bottomrule
\end{tabular}
\end{table}
 Thus, using $\eqref{pyrc}$ and Tables ~\ref{tabRC},\ref{tabCOP}, the p.m.f. for $(Y, X_1, X_2)$ is given in Table~\ref{tabYRC}.
\begin{table}[H]
\centering
\caption{Probability mass function of $(Y, X_1, X_2)$   } \label{tabYRC}
\begin{tabular}{cccc} \toprule
$y$ & $x_1$ & $x_2$ & $p(y, x_1, x_2)$ \\ \midrule
1 & 0 & 0 & $\lambda \, F(a_{00}-\tau)$ \\
0 & 0 & 0 &  $ \lambda \, (1 - F(a_{00}-\tau))$ \\
1 & $s_1$ & 0 & $\mu  \, F(a_{10} -\tau)$ \\
0 & $s_1$& 0 &  $ \mu  \, (1-F(a_{10} - \tau))$ \\
1 & 0 & $s_2$ & $\mu\,  F(a_{02} - \tau))$ \\
0 & 0 & $s_2$ &  $ \mu \, (1- F(a_{02} - \tau))$ \\
1 & $s_1$ & $s_2$ & $\lambda \, F(a_{12} - \tau)$ \\
0 & $s_1$ & $s_2$ &  $ \lambda\, (1- F(a_{12}- \tau))$ \\
\bottomrule
\end{tabular}
\end{table}

We define some ancillary functions of $s_1, s_2$ as follows
\begin{align}
\phi_1(s_2) & \equiv \Pr(Y =1, X_1 =0) = \lambda \, F(a_{00}-\tau) + \mu \, F(a_{02} - \tau) \label{p1}\\
\phi_2(s_1, s_2) &\equiv \Pr(Y =1, X_1 = s_1) = \mu \, F(a_{10}- \tau) + \lambda \, F(a_{12} - \tau) \label{p2}\\
\phi_3(s_1) & \equiv \Pr(Y =1, X_2 = 0) = \lambda \, F(a_{00}-\tau) + \mu \, F(a_{10} - \tau) \label{p3}\\
\phi_4(s_1, s_2) & \equiv \Pr(Y=1, X_2 = s_2) = \mu \, F( a_{02} - \tau) + \lambda \, F(a_{12} - \tau) \label{p4}
\end{align}
as well as the convex function
\begin{equation}
h(x) = -x  \log_2 (x) - (1 - x) \log_2 (1 - x), \quad (0 < x< 1). \label{h(x)}
\end{equation}
Using limiting arguments, we may deduce that $h(0) = h(1) =0$, so that $0 \leq h(x) \leq 1$. We note also that
\begin{equation}
\phi_1(s_2) + \phi_2(s_1, s_2) =  \phi_3(s_1) + \phi_4(s_1, s_2). \label{phi}
\end{equation}

\section*{Results}
\subsection*{Classical Information measures}
We now provide expressions for the information measures, with derivations provided   in the Appendix.
\begin{align}
hy(s_1, s_2) &\equiv H(Y) = h \left[ \phi_1(s_2) + \phi_2(s_1, s_2)  \nonumber  \right] \\
Iyx_1(s_1, s_2) &\equiv  I(Y ; X_1) = h \left[ \phi_1(s_2) + \phi_2(s_1, s_2)  \right] - \tfrac{1}{2} h \left[ 2  \phi_1(s_2) \right]  - \tfrac{1}{2} h \left[ 2  \phi_2(s_1, s_2) \right] \label{iyr}\\
Iyx_2[s_1, s_2] &\equiv I(Y ; X_2) = h \left[ \phi_3(s_1) + \phi_4(s_1, s_2)  \right] - \tfrac{1}{2} h \left[ 2 \phi_3(s_1) \right]  - \tfrac{1}{2} h \left[ 2 \phi_4(s_1, s_2) \right] \label{iyc}\\
Iyx_1x_2(s_1, s_1) &\equiv I(Y ; X_1, X_2) = h \left[ \phi_1(s_2) + \phi_2(s_1, s_2)  \right]  \nonumber\\
&\quad - \lambda \, h \left[ L(a_{00}-\tau) \right]  - \mu \, h \left[ L(a_{10} - \tau)\right]\ - \mu \, h \left[ L(a_{02} - \tau) \right]\ - \lambda \, h \left[ L(a_{12} - \tau \right] \label{iyrc}\\
Iyx_1Gx_2(s_1, s_2) &\equiv I(Y;X_1|X_2) = Iyx_1x_2(s_1, s_2) - Iyx_2(s_1, s_2) \label{iy1G2}  \\
Iyx_2Gx_1(s_1, s_2) &\equiv I(Y;X_2|X_1) = Iyx_1x_2(s_1, s_2) - Iyx_1(s_1, s_2) \label{iy2G1}\\
hyr(s_1, s_2) &\equiv H(Y)_\text{res} = hy(s_1, s_2) - Iyx_1x_2(s_1, s_2) \nonumber 
\end{align}

\subsection*{Exact Ibroja PIDs}

{\bf Theorem 1}

 For the transfer function defined in $\eqref{TF}$--$\eqref{ident}$ and the probability distribution defined in Table~\ref{tabYRC}, with $s_1 >0, s_2>0, -1< \rho <1$,  and the Ibroja PID:
 
  \noindent (a) When $a_{12} > a_{00}$,
  Unq$X_2$ =0 if $a_{10} > a_{02}$, (ii) Unq$X_1$ = 0 if $a_{10} < a_{02}$, (iii) Unq$X_1$ = Unq$X_2$ =0 if $a_{10} = a_{02}$. 
The exact Ibroja PID is defined by the following functions.

\begin{align}
\text{Unq}X_1(s_1, s_2) &=
\left\{
 \begin{array}  {lr}
 Iyx_1(s_1, s_2) -Iyx_2(s_1, s_2), & \text{if} \quad  a_{10} > a_{02} \\
                    0, & \text{if} \quad a_{10} \leq a_{02}
                         \end{array} 
                         \right.   \label{unqr}\\
\text{Unq}X_2(s_1, s_2) &= 
\left\{
\begin{array}{lr}
0, & \text{if} \quad a_{10} \geq a_{02} \\
                        Iyx_2(s_1, s_2) - Iyx_1(s_1, s_2), & \text{if} \quad a_{10}< a_{02}
                         \end{array} 
                         \right. \label{unqc}\\
\text{Shd}(s_1, s_2) &= 
\left\{
\begin{array}{lr}
Iyx_2(s_1, s_2), & \text{if} \quad a_{10} \geq a_{02} \\
                         Iyx_1(s_1, s_2), & \text{if} \quad a_{10}< a_{02}
                         \end{array} 
                         \right. \label{shd}\\
\text{Syn}(s_1, s_2) &= 
\left\{
\begin{array}{lr}
Iyx_2Gx_1(s_1, s_2), & \text{if} \quad a_{10} \geq a_{02} \\
                       Iyx_1Gx_2(s_1, s_2), & \text{if} \quad a_{10}< a_{02}
                         \end{array} 
                         \right. \label{syn}
\end{align}
(b) when $a_{12} \leq a_{00}$ and $a_{10} > a_{02}$, then Unq$X_1$ =0, and the exact Ibroja PID components are given by the following functions.

\begin{align}
\text{Unq}X_1(s_1, s_2) &= 0 \\
\text{Unq}X_2(s_1, s_2) &= Iyx_2(s_1, s_2) - Iyx_1(s_1, s_2) \\                        
\text{Shd}(s_1, s_2) &= Iyx_1(s_1, s_2) \\
 \text{Syn}(s_1, s_2) &= 
                       Iyx_1Gx_2(s_1, s_2)
                       \end{align}
\newline
\noindent {\bf Theorem 2}

 When $\rho = \pm 1$  in the probability distribution defined in Table~\ref{tabYRC} and the transfer function defined in $\eqref{TF}$--$\eqref{ident}$, with $s_1 >0, s_2>0$ and the Ibroja PID, the only nonzero PID component is the source shared information, ShdS, given by
 \begin{equation*}
 \text{ShdS}(s_1, s_2) = \begin{cases}  \, h( \tfrac{1}{2} ( F(a_{00}-\tau) + F(a_{12} -\tau))) - \tfrac{1}{2} h(F(a_{00}-\tau))    - \tfrac{1}{2} h(F(a_{12} -\tau)),\,\,  \text{if}\quad  \rho = +1, \\
   \, h( \tfrac{1}{2} ( F(a_{10}-\tau) + F( a_{02} -\tau)) - \tfrac{1}{2} h(F(a_{10}-\tau)))    - \tfrac{1}{2} h(F(a_{02} -\tau)),  \,\,\, \text{if}\quad \rho = -1.     \end{cases}
 \end{equation*}
 Proofs of these theorems are provided in the Appendix.
 
 \subsection*{Some general transfer functions}
 We consider some general additive and multiplicative forms for the activation function $a(x_1, x_2)$ in (2),  based partly on \citet{BAA}, \citet{RAS} and \citet{KP}. Let the functions $f$, $g$ map 
 $\mathbb{R}_{\geq 0}$ to $\mathbb{R}_{\geq 0}$, function $g_1$ map $\mathbb{R}_{\geq 0}$ to the set of positive real numbers $\mathbb{R}^+$, and function $h$ map  $\mathbb{R}_{\geq 0} \times \mathbb{R}_{\geq 0}$ to $\mathbb{R}_{\geq 0}$, and suppose that each of these functions is continuous.
 
 Suppose that these functions have the following properties.
 \begin{equation}
 f(0) =0,   f(x_1) >0\,\, (x_1 >0);  g(0) =0, \,\, g(x_2) >0,\, (x_2 >0); \quad g_1(x_2) >0, \,\,(x_1, x_2 \geq 0) \label{cond1}
 \end{equation}
 \begin{equation}
  h(0, x_2) = h(x_1,0) = 0, \,\,h(x_1, x_2)>0 \quad (x_1 >0, x_2 >0) \label{cond2}
 \end{equation}
 We consider the following general activation functions. Given inputs, the value of each one is then passed through a sigmoidal function to give the value of the transfer function.
 \begin{align}
 a(x_1, x_2) &= f(x_1)g(x_2). \label{a1} \\
 a(x_1, x_2) &= f(x_1) g_1(x_2) \label{a2}\\
 a(x_1, x_2) &= f(x_1)g_1(x_1x_2) \label{a3}\\
 a(x_1, x_2) &= f(x_1) + g(x_2) \label{a4}\\
 a(x_1, x_2) &= f(x_1) -  g(x_2) \label{a5}\\
 a(x_1, x_2) &= f(x_1) + g(x_2) + h(x_1, x_2) \label{a6}
 \end{align}
 By application of Theorem 1, we obtain exact Ibroja PIDs for each of these general forms of activation function.
 
 For the activation function in \eqref{a1}, and using \eqref{cond1}, \eqref{cond2}, we have
 \begin{equation}
 a_{00} = a_{10} = a_{02} = 0, a_{12} = f(s_1)g(s_2)>0.
 \end{equation}
 Since $a_{12} > a_{00}$ and $a_{10} = a_{02}$ we may read the Ibroja PID from (15)--(18) and from (10), (11) and (13),(14). In particular, we see that there is no transmission of unique information because each of the two inputs has no unique information about the output. 
 
 For the activation function in \eqref{a2}, and using \eqref{cond1}, \eqref{cond2}, we have
 \begin{equation}
 a_{00} = 0, a_{10} = f(s_1) g_1(0) >0, a_{02} = 0, a_{12} = f(s_1)g_1(s_2)>0.
 \end{equation}
 Since $a_{12} > a_{00}$ and $a_{10} > a_{02}$ we may read the Ibroja PID from (15)--(18) and from (10), (11) and (13),(14). In particular, we see that input $X_2$ has no unique information about the output.
 
 For the activation function in \eqref{a3}, and using \eqref{cond1}, \eqref{cond2}, we have
 \begin{equation}
 a_{00} = 0, a_{10} = f(s_1) g_1(0) >0, a_{02} = 0, a_{12} = f(s_1)g_1(s_1s_2)>0.
 \end{equation}
 Since $a_{12} > a_{00}$ and $a_{10} > a_{02}$ we may read the Ibroja PID from (15)--(18) and from (10), (11) and (13),(14). In particular, we see also here that input $X_2$ has no unique information about the output.
 Of the four modulatory functions considered in \citet{KP},  one is of the form in \eqref{a2}  while the other three have the form of \eqref{a3}. 
 
 These activation functions are of multiplicative form but for the additive form in \eqref{a4}, and using \eqref{cond1}, \eqref{cond2}, we have
 \begin{equation}
 a_{00} = 0, a_{10} = f(s_1)  , a_{02} =  g(s_2), a_{12} = f(s_1) + g(s_2)>0.
 \end{equation}
 Since $a_{12} > a_{00}$ we may read the Ibroja PID from (15)--(18) and from (10), (11) and (13),(14), depending on whether $f(s_1)$ is greater than, less than, or equal to  $g(s_2)$. In the first case, input $X_2$ has no unique information about the output, in the second case input $X_1$ has no unique information about the output, and in the third case both unique information terms are equal to zero.
 
 The activation function in \eqref{a5} has a subtractive form with $X_1$ driving and the $X_2$ value being an integrated value based on the outputs of neighbouring cells. We have using \eqref{cond1}, \eqref{cond2}
 \begin{equation}
  a_{00} = 0, a_{10} = f(s_1)  , a_{02} =  -g(s_2), a_{12} = f(s_1) - g(s_2).
 \end{equation}
 
When $f(s_1) > g(s_2)$ then $a_{12} > a_{00}$. Also $a_{10} > a_{02}$ since $f,g$ are positive given the positive input strengths $s_1, s_2$. From Theorem 1(a) we see that Unq$X_2=0$ and we may read the Ibroja components from from (15)--(18) and from (10), (11) and (13),(14). When $f(s_1) < g(s_2)$ then $a_{12} < a_{00}$. Also $a_{10} > a_{02}$ since $f,g$ are positive given the positive input strengths $s_1, s_2$. From Theorem 1(b) we see that Unq$X_1=0$ and we may read the Ibroja components from (19)--(22). When $f(s_1) = g(s_2)$ both unique components are equal to zero.

The activation function in \eqref{a6} is more general. We have using \eqref{cond1}, \eqref{cond2}
\begin{equation}
  a_{00} = 0, a_{10} = f(s_1)  , a_{02} =  g(s_2), a_{12} = f(s_1) + g(s_2) +h(s_1, s_2) >0.
 \end{equation}
 As with the additive activation function \eqref{a4}, since $a_{12} > a_{00}$ we may read the Ibroja PID from (15)--(18) and from (10), (11) and (13),(14), depending on whether whether $f(s_1)$ is greater than, less than, or equal to $g(s_2)$. In the first case, input $X_2$
has no unique information about the output, in the second case input $X_1$ has no unique information about the output, and in the third case both unique information terms are equal to zero.

  We now consider some particular applications of Theorem 1. We take the function $F$ to be the cumulative distribution function of the logistic distribution which is a strictly increasing continuous function. We also take $\rho =0$, which from Table 1 means that the inputs $X_1$ and $X_2$ are independent. This means that the source-shared PID component is equal to zero, so all transmitted shared information is mechanistic and Shd = ShdM.  We consider values of $s_1$ and $s_2$ in the range [0.1, 10], and take the threshold parameter, $\tau$, to be equal to 5. We are interested in how the various combinations of the input signal strengths, $s_1, s_2$, affect the transmission of information within the system. We consider the PID components for the trivariate probability distribution in Table~\ref{tabYRC} for given values of $s_1, s_2$, which we refer to as an eight-point distribution based on the $(s_1, s_2)$ values: $(0,0), (s_1, 0), (0, s_2)$ and $(s_1, s_2)$, and also the 0/1 values for the output. This is repeated for  values of $s_1, s_2$ over the square $[0.1, 10] \times [0.1, 10]$.

 \subsection*{Application: A modulatory transfer function}
Taking $a(x_1, x_2) = x_1 + x_1  x_2= x_1(1 +x_2)$, we consider the transfer function
\begin{equation}
T_1(x_1, x_2) = F(x_1 + x_1 x_2 - \tau). \label{T1}
\end{equation}
This activation function has the multiplicative form defined in \eqref{a2}, with $f(x_1) =x_1$ and $g_1(x_2) = 1 + x_2$. We take input $X_1$ to be providing the drive, and  input $X_2$ to be the context. Then from (5) it follows that
\begin{equation*}
a_{00} =0, \quad a_{10} = s_1, \quad a_{02} = 0, \quad a_{12} = s_1 + s_1 s_2,
\end{equation*}
and recall that $s_1 >0, s_2 >0.$ Clearly $a_{12} > a_{00}$ so we can use Theorem 1(a). Also, $a_{10} > a_{02}$ and so we can read the PID component functions  from $\eqref{unqr}$--$\eqref{syn}$ in Theorem 1, and from $\eqref{iyr}$,$\eqref{iyc}$ and  $\eqref{iy1G2}$,$\eqref{iy2G1}$.

\begin{figure}[H]
\centering
\includegraphics[scale =0.6]{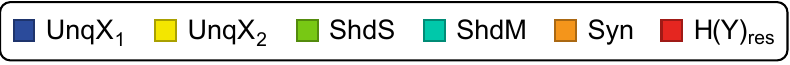}

\begin{tabular}{cr}
\includegraphics[scale=0.5]{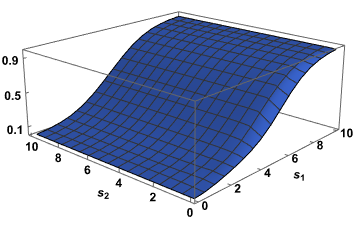} & \includegraphics[scale=0.5]{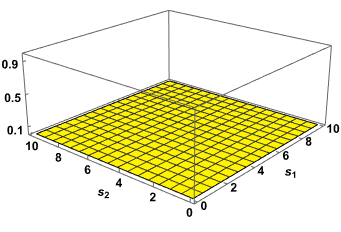}\\
\includegraphics[scale=0.5]{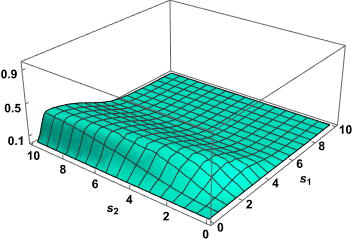} & \includegraphics[scale=0.5]{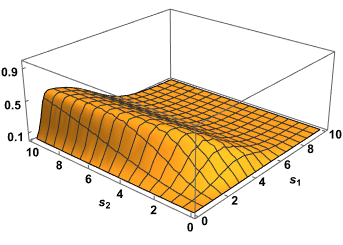} \\
\end{tabular}
\caption{Surfaces of the normalised PID components using the modulatory transfer function $T_1$, defined in $\eqref{T1}$} \label{Fig1_surf}
\end{figure}
\begin{figure}[H]
\centering
\includegraphics[scale =0.5]{FIGURES/leg_mca.pdf}

\begin{tabular}{cc}
\includegraphics[scale =0.5]{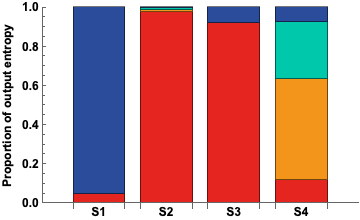}
& \includegraphics[scale=0.5]{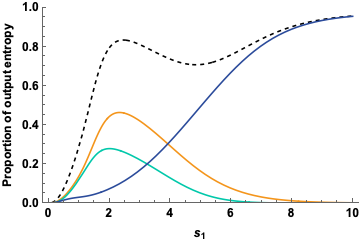}  \\
\end{tabular}
\caption{For transfer function $T_1$ in $\eqref{T1}$, normalised Ibroja PIDs for strength scenarios S\textsubscript{1} $(s_1 = 10, s_2 = 0.1)$, S\textsubscript{2} $(s_1 = 0.1, s_2=  10)$, S\textsubscript{3} $(s_1 = 4, s_2=  1)$ , S\textsubscript{4} $(s_1=  4, s_2 = 3)$;  The line plot shows the values of the normalised PID components as the $X_1$ input is increased from 0.1  to 10, and $X_2$ is fixed at 3. The dashed line traces the normalised joint mutual information. } \label{Fig2_blp}
\end{figure}

In Fig.~\ref{Fig1_surf}, the plot of Unq$X_2$ illustrates the fact that this component is equal to zero for all combinations of the input signal strengths, so $X_2$ holds no unique information about the output. As the strength of input $X_1$ is increased from 0.1 to 10.0, more and more  unique information held by  $X_1$ is transmitted; this component is the dominant one for large values of $s_1$. When the strength of $X_1$ is low (0-4), we see that the synergy component rises and falls, irrespective of the strength of $X_2$. 

The surface for mechanistic shared information lies below that of the synergy (as expected from \citet[Eq. (10)]{PPCP}) and follows a similar pattern. We also see that for large values of $s_1$ and $s_2$ both the synergy and mechanistic shared components are close to zero; thus there is no evidence of signal integration for transfer function $T_1$.

The barplot for Scenario S\textsubscript{1} in Fig.~\ref{Fig2_blp} indicates that, when $X_1$ is strong and $X_2$ is close to zero, almost all of the information is transmitted as being unique information held by the $X_1$ input. This indicates that condition CM1 is satisfied.   The barplot for scenario S\textsubscript{2} comprises almost entirely of residual information, with almost no information being transmitted. Thus even when the context $X_2$ is strong with almost zero drive $X_1$ there is insufficient information transmitted for an event to occur, supporting CM2. In Scenario S\textsubscript{3}, some unique information held by the input $X_1$  is transmitted but input $X_2$ has no unique information  about the output, thus supporting condition CM3. The PID for Scenario S\textsubscript{4} indicates that  when the $X_1$ input is weak an increase in the $X_2$ input results in an increase in the transmission of information in the form of synergistic and  mechanistic shared information as well as some unique information from input $X_1$. Thus condition CM4 is satisfied. The line plot in Fig.~\ref{Fig2_blp} further illustrates the  comments on the surfaces in Fig.~\ref{Fig1_surf} regarding the rise and fall of the information transmitted in the form of mechanistic shared and synergistic information, suggesting that the condition for signal integration is not present. Hence, we conclude that the transfer function $T_1$ is purely modulatory.

An alternative approach here would be to swap $X_1$ with $X_2$ and have $X_2$ driving and being modulated by $X_1$ by taking $a(x_1, x_2) = x_2 + x_2  x_1= x_2(1 +x_1)$ in  the transfer function
\begin{equation}
T_2(x_1, x_2) = F(x_2 + x_2 x_1 - \tau).
\end{equation}
Clearly, this also has the  multiplicative form \eqref{a2}. Then from (5) it follows that
\begin{equation}
a_{00} =0, \quad a_{10} = 0, \quad a_{02} = s_2, \quad a_{12} = s_2 + s_2 s_1,
\end{equation}
and recall that $s_1 >0, s_2 >0.$ Clearly $a_{12} > a_{00}$ so we can use Theorem 1(a). Also, $a_{10} <  a_{02}$ and so we can read the PID component functions from $\eqref{unqr}$--$\eqref{syn}$ in Theorem 1(a), and from $\eqref{iyr}$,$\eqref{iyc}$ and  $\eqref{iy1G2}$,$\eqref{iy2G1}$. We find that Unq$X_1$ = 0 rather than Unq$X_2$ =0. We omit the plots in this case since they are  essentially the same as those in Figs.~\ref{Fig1_surf} and~\ref{Fig2_blp}, after swapping $s_1$ and $s_2$ and the colours of the unique information components. Hence $T_2$ is also a modulatory transfer function through which $X_1$ modulates $X_2$, when $X_2$ is weak.

\subsection*{Application: Another  modulatory transfer function}
Taking $a(x_1, x_2) = x_1(1 + \exp{(0.1 x_1 x_2}))$, we consider the transfer function
\begin{equation}
T_3(x_1, x_2) = F(x_1(1 + \exp{(0.1 x_1 x_2})) - \tau). \label{T3}
\end{equation}
This activation function has the multiplicative form defined in \eqref{a3}, with $f(x_1) =x_1$ and $g_1(x_1x_2) = 1 + \exp{(0.1 x_1 x_2})$. We take input $X_1$ to be providing the drive, and  input $X_2$ to be the context. The results in this case are very close to those for transfer function $T_1$ and so the same conclusions apply; we omit the details.

\subsection*{Application: An integrative transfer function}
Taking $a(x_1, x_2) = x_1  x_2$, we consider the transfer function
\begin{equation}
T_4(x_1, x_2) = F(x_1 x_2 - \tau). \label{T4}
\end{equation}
The activation function has the multiplicative form defined in \eqref{a1}. We take input $X_1$ to be providing the drive, and  input $X_2$ to be the context although, since the transfer function is symmetric, we could also swap the roles of these inputs.  Then from (5) it follows that
\begin{equation*}
a_{00} =0, \quad a_{10} = 0, \quad a_{02} = 0, \quad a_{12} =  s_1 s_2,
\end{equation*}
and recall that $s_1 >0, s_2 >0.$ Clearly $a_{12} > a_{00}$ so we can use Theorem 1(a). Also, $a_{10} = a_{02}$ and so we can read the PID component functions from $\eqref{unqr}$--$\eqref{syn}$ in Theorem 1(a), and from $\eqref{iyr}$,$\eqref{iyc}$ and  $\eqref{iy1G2}$,$\eqref{iy2G1}$.

\begin{figure}[H]
\centering
\includegraphics[scale =0.6]{FIGURES/leg_mca.pdf}

\begin{tabular}{cr}
\includegraphics[scale=0.42]{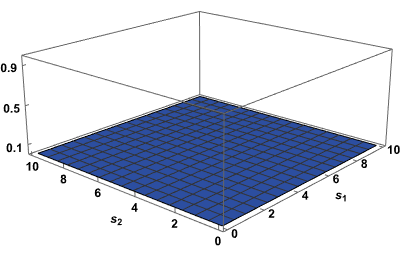} & \includegraphics[scale=0.5]{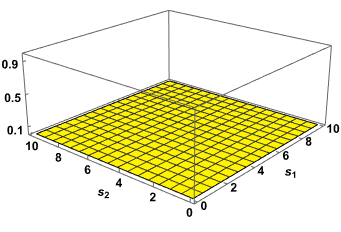}\\
\includegraphics[scale=0.5]{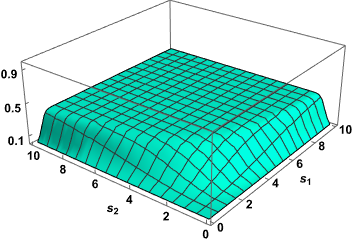} & \includegraphics[scale=0.5]{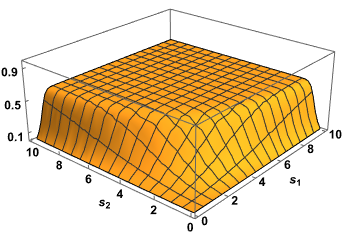} \\
\end{tabular}
\caption{Surfaces of the normalised PID components for the integrative transfer function $T_4$, defined in $\eqref{T4}$.} \label{Fig3_surf}
\end{figure}
\begin{figure}[H]
\centering

\begin{tabular}{cc}
\includegraphics[scale =0.4]{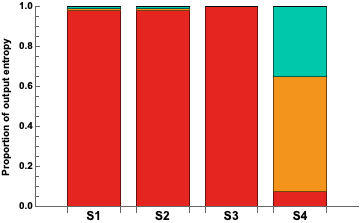}
& \includegraphics[scale=0.4]{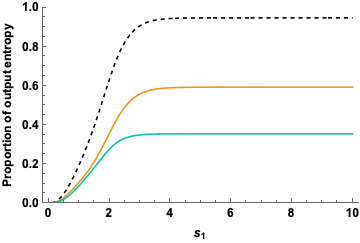}  \\
\end{tabular}
\caption{For transfer function $T_4$ in $\eqref{T4}$, normalised Ibroja PIDs for strength scenarios S\textsubscript{1} $(s_1= 10, s_2 = 0.1)$, S\textsubscript{2} $(s_1 = 0.1, s_2=  10)$ , S\textsubscript{3} $(s_1 = 2, s_2=  0.1)$ , S\textsubscript{4} $(s_1 =  2, s_2 = 5)$;  A line plot showing the values of the normalised PID components as the $X_1$ input is increased from 0.1  to 10, and  $X_2$ is  fixed at 3. The dashed line traces the normalised joint mutual information.  } \label{Fig4_blp}
\end{figure}

We see from Theorem 1 and the plots in Fig.~\ref{Fig3_surf} that both components of unique information are equal to zero for all combinations of the strengths of the inputs. Thus the joint mutual information comprises only  mechanistic shared and synergistic information, which have symmetric surfaces; it is clear that when both signal strengths are high all the information is transmitted via the mechanistic shared and synergistic components, while no unique information at all is transmitted. This provides clear evidence for the integration of the input signals, $X_1, X_2$.

The bar plots in Fig.~\ref{Fig4_blp} show that for scenarios S\textsubscript{1}--S\textsubscript{3} almost no information is transmitted, while in scenario S\textsubscript{4} almost all of the information available is transmitted in the form of mechanistic shared and also synergy. Hence there is no support in this case for conditions CM1--CM4. The line plot in Fig.~\ref{Fig4_blp} shows no rise and fall in mechanistic shared and synergistic information, thus reflecting the nature of the corresponding surface plots,   but rather a rise to a maximum and level off. Thus, we conclude that information processing with transfer function $T_4$ is entirely integrative.

\subsection*{Application: Modulation and integration when inputs are combined additively}
Taking $a(x_1, x_2) = x_1 +   x_2$, we consider the transfer function
\begin{equation}
T_5(x_1, x_2) = F(x_1 +  x_2 - \tau). \label{T5}
\end{equation}
This activation function has the additive form defined in \eqref{a4}, with $f, g$ taken to be the identity mapping. Then from (5) it follows that
\begin{equation*}
a_{00} =0, \quad a_{10} = s_1, \quad a_{02} = s_2, \quad a_{12} = s_1 +  s_2,
\end{equation*}
and recall that $s_1 >0, s_2 >0.$ Clearly $a_{12} > a_{00}$ so we can use Theorem 1(a). There are three cases to consider: $s_1 > s_2, s_1 < s_2$ and $s_1 = s_2$,   and we can read the PID component functions in each case from from $\eqref{unqr}$--$\eqref{syn}$ in Theorem 1(a), and from $\eqref{iyr}$,$\eqref{iyc}$ and  $\eqref{iy1G2}$,$\eqref{iy2G1}$.
\begin{figure}[H]
\centering
\includegraphics[scale =0.6]{FIGURES/leg_mca.pdf}

\begin{tabular}{cr}
\includegraphics[scale=0.5]{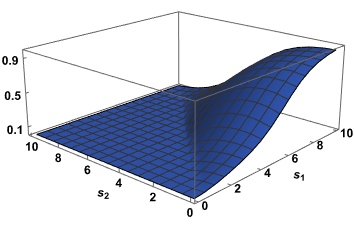} & \includegraphics[scale=0.5]{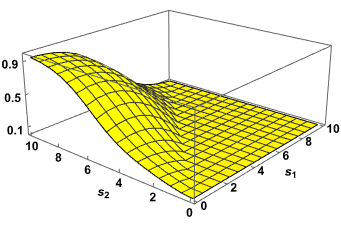}\\
\includegraphics[scale=0.5]{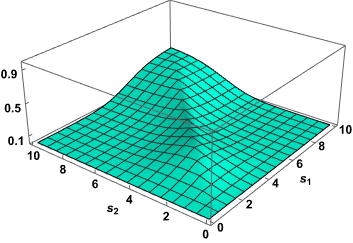} & \includegraphics[scale=0.5]{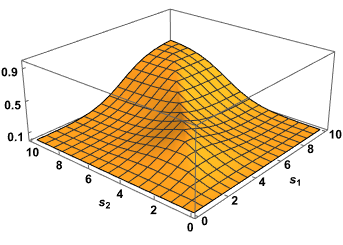} \\
\end{tabular}
\caption{Surfaces of the PID components for the  transfer function $T_5$ in $\eqref{T5}$, in which the inputs are combined additively} \label{Fig5_surf}
\end{figure}

In Fig.~\ref{Fig5_surf}, the Unq$X_1$ and Unq$X_2$ surfaces show a zero triangular region, as proved in Theorem 1, and these two surfaces are mirror images of each other. When the $X_2$ input is low (0-2) the value of Unq$X_1$ rises steadily, and vice-versa for low $X_1$ input and unq$X_2$. The shared mechanistic and synergy surfaces are symmetric about the line $s_1 = s_2$, and both components generally increase as $s_1$ and $s_2$ are increased along the line $s_1 = s_2$, with the ShdM component less than or equal to the Syn component (Equation (10), Pica et al., 2017). When the $X_2$ input lies between 2 and 4, we notice that the synergy component rises and then falls as the $X_1$ input is increased towards 10. Similarly, when the $X_1$ input lies between 2 and 4 the synergy component steadily increases as the $X_2$ input is increased towards 10. We now check the conditions for contextual modulation.

\begin{figure}[H]
\centering
\includegraphics[scale =0.5]{FIGURES/leg_mca.pdf}

\begin{tabular}{cc}
\includegraphics[scale =0.45]{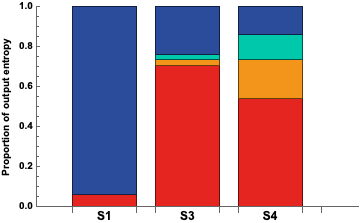}
& \includegraphics[scale=0.45]{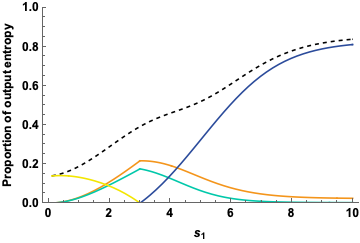}  \\
\end{tabular}
\caption{For the transfer function $T_5, (s_1 \geq s_2)$ in $\eqref{T5}$, normalised Ibroja PIDs for strength scenarios S\textsubscript{1} $(s_1 = 10, s_2 = 0.1)$, S\textsubscript{3} $(s_1 = 4, s_2=  1)$, S\textsubscript{4} $(s_1 =  4, s_2 = 3)$;  Line plots showing the values of the normalised PID components as the $X_1$ input is increased from 0.1  to 10, and $X_2$ is  fixed at 3. The dashed line traces the normalised joint mutual information. } \label{Fig6_blp}
\end{figure}

\subsection*{Does the  input $X_2$ modulate the  input  $X_1$($\mathbf {s_1 > s_2}$)?}
Consider the bar plots in Fig.~\ref{Fig6_blp}. 
The PID for Scenario S\textsubscript{1} indicates that almost all of the information is transmitted as being unique information about the output conveyed by the $X_1$ input, thus indicating that condition CM1 is satisfied.  Scenario S\textsubscript{2} cannot be checked. In Scenario S\textsubscript{3}, some unique information held by the $X_1$ input is transmitted but  the $X_2$ input possesses no unique information about the output, thus supporting condition CM3. The PID for Scenario S\textsubscript{4} indicates that  when the $X_1$ input is weak an increase in the $X_2$ input results in an increase in the transmission of information in the form of synergistic, mechanistic shared information as well as unique information provided by the $X_1$ input.

The line plot in Fig.~\ref{Fig6_blp} show the profiles of the information components as the $X_1$ input is increased from 1 to 10.  It is clear  that the transmission of synergistic and shared information rises and then falls as the level of $X_1$ is increased, reflecting the comments on the associated surfaces in Fig.\ref{Fig5_surf}. We notice also that when the values of these PID components starts to fall the transmission of unique information conveyed by $X_1$ rises and eventually is equal  or close to the joint mutual information. On the contrary, the unique information possessed by the $X_2$ input is typically close to zero. Thus conditions CM1, CM3-4 are  satisfied and this supports modulation of the  input $X_1$ by input $X_2$ when $X_1$ is weak.

\subsection*{Does the input $X_1$ modulate the  input $X_2$ ($\mathbf{s_1 <  s_2}$)?}
\begin{figure}[H]
\centering
\includegraphics[scale =0.5]{FIGURES/leg_mca.pdf}

\begin{tabular}{cc}
\includegraphics[scale =0.45]{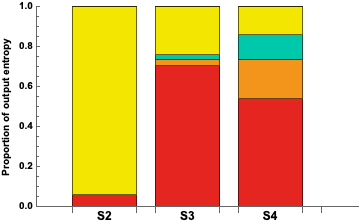}
& \includegraphics[scale=0.45]{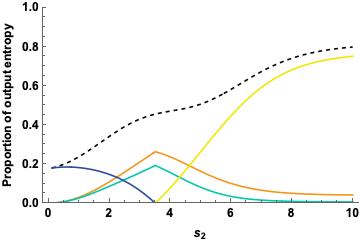}  \\
\end{tabular}
\caption{For the transfer function $T_5, (s_1 \leq s_2)$ in $\eqref{T5}$, normalised Ibroja PIDs for strength scenarios S\textsubscript{2} $(s_1 = 0.1, s_2 = 10)$, S\textsubscript{3} $(s_1 = 1, s_2=  4)$, S\textsubscript{4} $(s_1 =  3, s_2 = 4)$; Line plots showing the values of the PID components as the $X_2$ input is increased from 0.1  to 10, and  $X_1$ is  fixed at 3.5. The dashed line traces the normalised joint mutual information. } \label{Fig7_blp}
\end{figure}

The plots in Fig.~\ref{Fig7_blp} are essentially the mirror images of those in Fig.~\ref{Fig6_blp}. Thus, conditions CM2--CM4 are satisfied here and this supports  modulation of the $X_2$ input by $X_1$ when $X_2$ is weak. We see that, even with an additive transfer function, if one input is stronger than the other then the weaker input provides contextual modulation for the other in terms of information transmission.

\subsection*{Integration of the inputs, $\mathbf{X_1, X_2}$ }

The surface plots of the mechanistic shared and synergy components in Fig.~\ref{Fig5_surf} suggest for high values of $s_1$ and $s_2$, when $s_1$ and $s_2$ are both increased, that the synergy and shared information increase and start to level off. 
\begin{figure}[ht]
\centering
\includegraphics[scale =0.5]{FIGURES/leg_mca.pdf}
\begin{tabular}{cc}
 \includegraphics[scale =0.5]{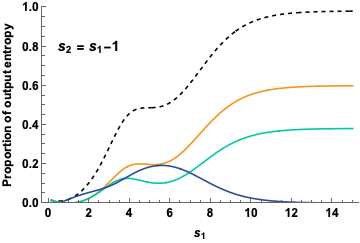}& \includegraphics[scale=0.5]{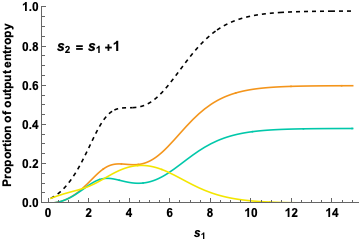}  \\
\end{tabular}
\includegraphics[scale=0.5]{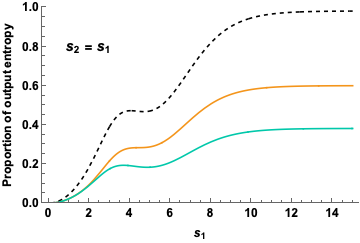} 
\caption{Plots of the normalised PID components along the lines: $s_2 = s_1-1$, $s_2 = s_1+ 1$, $s_2 = s_1$ using transfer function $T_5$ in $\eqref{T5}$. The dashed line traces the normalised joint mutual information. } \label{Fig8}
\end{figure}

\begin{figure}[H]
\centering
\includegraphics[scale=0.5]{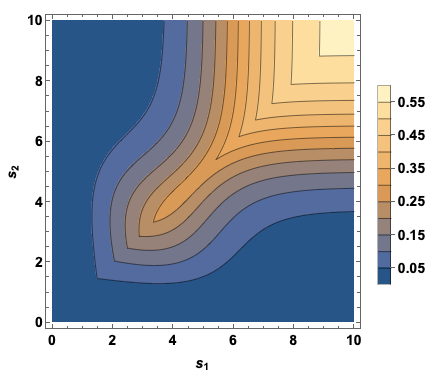} 
\caption{For transfer function $T_5$ in $\eqref{T5}$, a contour plot of the synergy component } \label{Fig9}
\end{figure}

The plots in Fig.~\ref{Fig8} show that this is the case along the line of $s_1$-$s_2$ equality and nearby parallel lines. What is also of interest is that there is a brief levelling off, or slight decrease, in the synergy values for lower values of $s_1$ and $s_2$. Further insight is provided by the contour plot of Syn in Fig.~\ref{Fig9}. When $s_1$ and/or $s_2$ are/is low (2-4), fixing say $s_1=3$ and letting $s_2$ increase to 10 leads to a rise and fall in synergy; similarly if $s_2$ is fixed in the range [2,4] and $s_1$ increased to 10. On the other hand, when $s_1$ and $s_2$ are both high, say larger than 5, examination of the synergy profile along lines such as $s_1 =6$ or $s_2 =7$ shows that the synergy rises and levels off; also increasing $s_1$ and $s_2$ simultaneously on or near the line $s_1 =s_2$ leads to  increasing synergy towards a saturating level. This saturation is clearer in the line plots in Fig.~\ref{Fig8}. Thus, there is a suggestion of three possibilities here: (i) low $s_1$ with unrestricted $s_2$  in which $X_1$ can modulate $X_2$, (ii)  and low $s_2$  with unrestricted $s_1$ in which $X_2$ can modulate $X_1$, and (iii) high $s_1$ and $s_2$ where there is integration of both inputs leading to increasing synergy and shared information coupled with decreasing  or little unique information of either kind.

\subsection*{Application: Modulation and integration when the inputs are combined subtractively}
Taking $a(x_1, x_2) = x_1 -   x_2$, we consider the transfer function
\begin{equation}
T_6(x_1, x_2) = F(x_1 -  x_2 - \tau). \label{T6}
\end{equation}
This activation function has the subtractive form defined in \eqref{a5}, with $f,g$ taken to be the identity mapping.Then from (5) it follows that
\begin{equation*}
a_{00} =0, \quad a_{10} = s_1, \quad a_{02} = - s_2, \quad a_{12} = s_1 -  s_2,
\end{equation*}
and recall that $s_1 >0, s_2 >0.$ Clearly $a_{10} > a_{02}$ so we can use Theorem 1(a). There are two cases to consider: when $s_1 > s_2,$  Unq$X_2$ =0, and we can read the PID component functions in each case from from $\eqref{unqr}$--$\eqref{syn}$ in Theorem 1(a), and from $\eqref{iyr}$,$\eqref{iyc}$ and  $\eqref{iy1G2}$,$\eqref{iy2G1}$. When $s_1 < s_2$, Unq$X_1$ =0, and we can read the PID component functions from (19)-(22) in Theorem 1(b) along with $\eqref{iyr}$,$\eqref{iyc}$ and  $\eqref{iy1G2}$,$\eqref{iy2G1}$.

\begin{figure}[H]
\centering
\includegraphics[scale =0.6]{FIGURES/leg_mca.pdf}
\begin{tabular}{cr}
\includegraphics[scale=0.5]{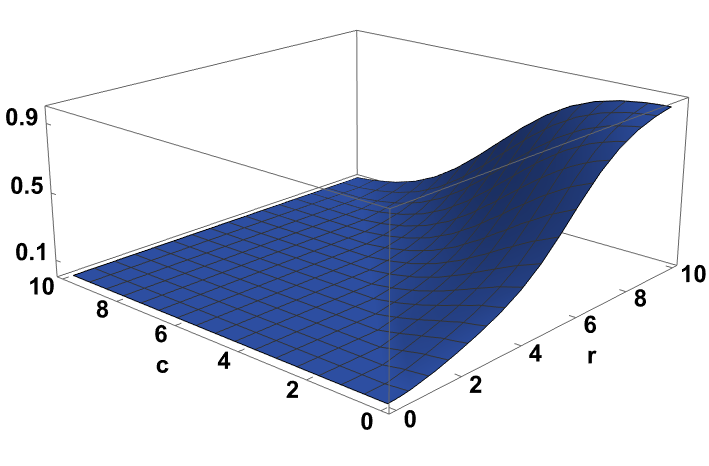} & \includegraphics[scale=0.5]{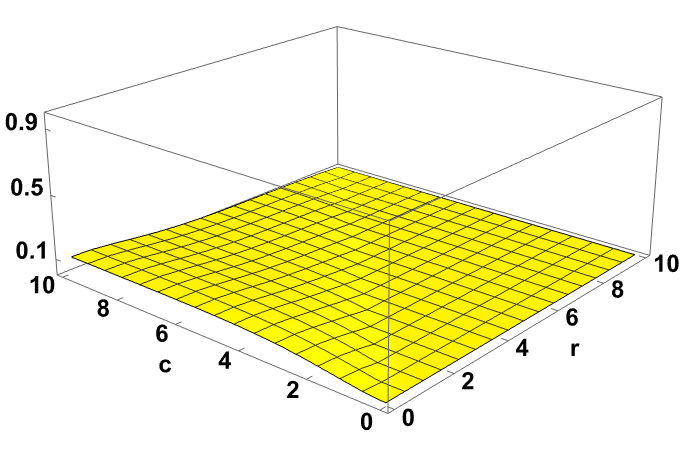}\\
\includegraphics[scale=0.5]{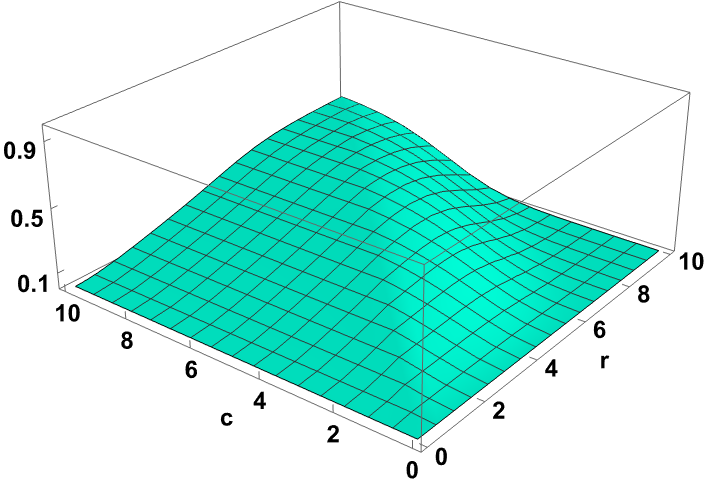} & \includegraphics[scale=0.5]{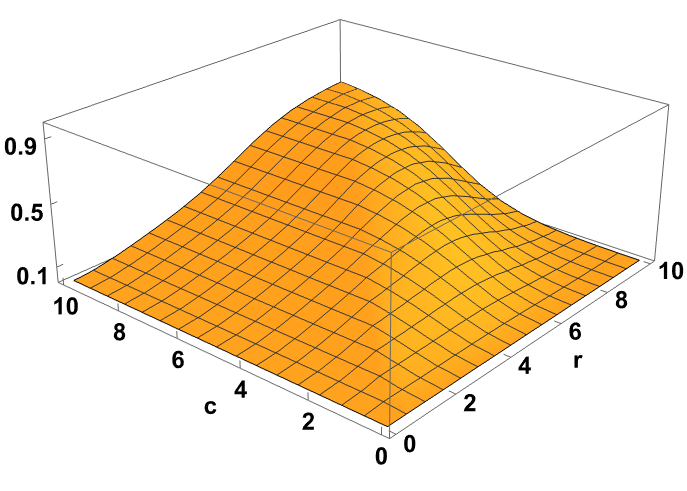} \\
\end{tabular}
\caption{Surfaces of the PID components for the  transfer function $T_6$ in $\eqref{T6}$, in which the inputs are combined subtractively} \label{Fig10_surf}
\end{figure}

In Fig.~\ref{Fig10_surf}, we notice the mentioned zero triangular regions for Unq$X_1$ and Unq$X_2$ and that the two functions are markedly different. When the $X_2$ input is low (0-2) the value of Unq$X_1$ rises steadily towards a high value whereas when the $X_1$ input is low (0-2) the value of Unq$X_2$ barely rises at all. The ShdM and Syn surfaces have substantial values for large signal strengths. When the $X_2$ input is between 2 and 4 we see that the ShdM and Syn components rise and fall as $X_1$ increases towards 10; conversely, when the $X_1$ input is between 2 and 4 these components rise but appear to stay constant as $X_2$ increases towards 10.

\subsection*{Does the  input $X_2$ modulate the  input  $X_1$ ($\mathbf {s_1 >  s_2}$)?}
\begin{figure}[H]
\centering
\includegraphics[scale =0.5]{FIGURES/leg_mca.pdf}
\begin{tabular}{cc}
\includegraphics[scale =0.45]{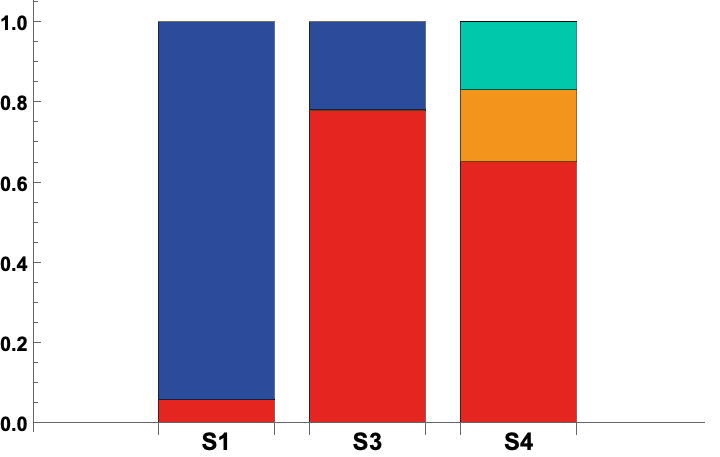}
 & \includegraphics[scale=0.45]{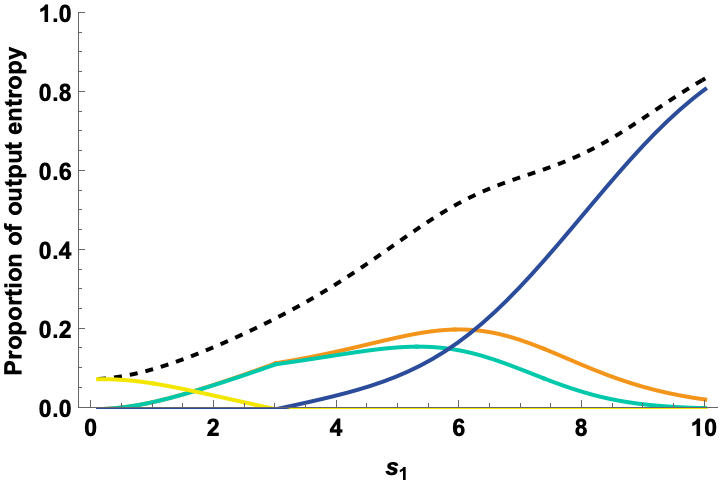}  \\
\end{tabular}
\caption{For the transfer function $T_6, (s_1 \geq s_2)$ in $\eqref{T6}$, normalised Ibroja PIDs for strength scenarios S\textsubscript{1} $(s_1 = 10, s_2 = 0.1)$, S\textsubscript{3} $(s_1 = 4, s_2=  1)$, S\textsubscript{4} $(s_1 =  4, s_2 = 3)$;  Line plots showing the values of the normalised PID components as the $X_1$ input is increased from 0.1  to 10, and $X_2$ is  fixed at 3. The dashed line traces the normalised joint mutual information. } \label{Fig11_blp}
\end{figure}

The bar plots in Fig.~\ref{Fig11_blp} suggest an affirmative answer, since conditions CM1, CM3-4 appear to be satisfied since they are similar to those in Fig.~\ref{Fig6_blp}. The line plot illustrates the rise and fall of the ShdM and Syn components when $X_1$ is weak; also the Unq$X_1$ component rises for larger values of $X_1$ towards the value of the JMI.
\subsection*{Does the  input $X_1$ modulate the  input  $X_2$ ($\mathbf {s_1 <  s_2}$)?}
\begin{figure}[H]
\centering
\includegraphics[scale =0.5]{FIGURES/leg_mca.pdf}
\begin{tabular}{cc}
\includegraphics[scale =0.45]{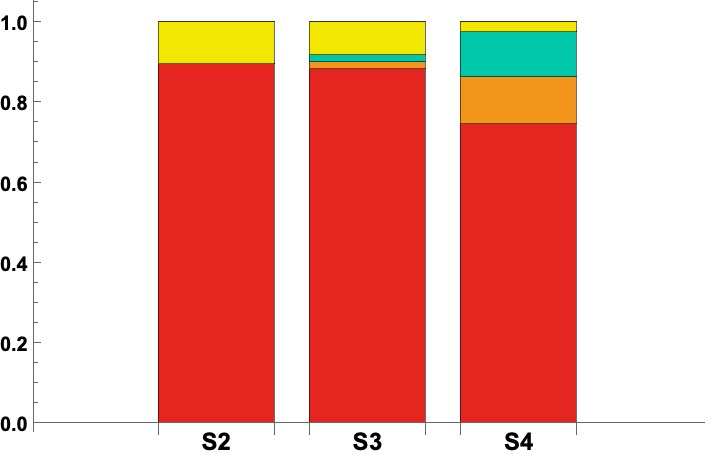}
& \includegraphics[scale=0.45]{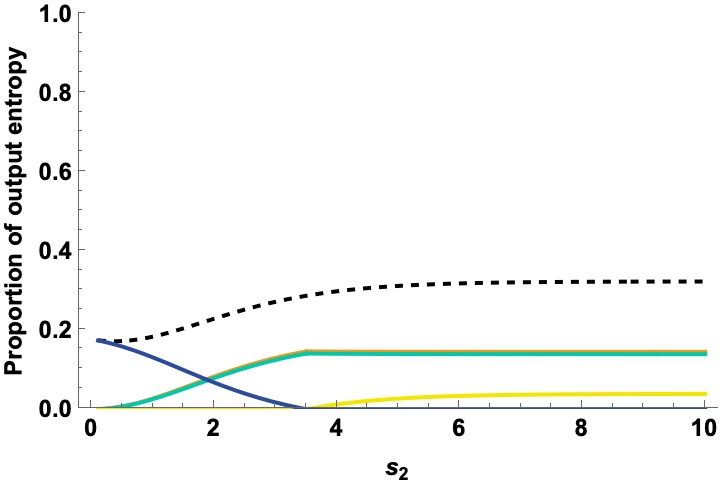}  \\
\end{tabular}
\caption{For the transfer function $T_6, (s_1 \leq s_2)$ in $\eqref{T6}$, normalised Ibroja PIDs for strength scenarios S\textsubscript{2} $(s_1 = 0.1, s_2 = 10)$, S\textsubscript{3} $(s_1 = 1, s_2=  4)$, S\textsubscript{4} $(s_1 =  3, s_2 = 4)$; Line plots showing the values of the PID components as the $X_2$ input is increased from 0.1  to 10, for $X_1$  fixed at 3.5. The dashed line traces the normalised joint mutual information. } \label{Fig12_blp}
\end{figure}

Consider the plots in Fig.~\ref{Fig12_blp}. In Scenario S2 with strong $X_2$ and very weak $X_1$ very little unique information is proved by $X_2$ and only a small proportion of the available bandwidth is used in transmitting this information. Therefore  $X_2$ is not sufficient with regard to being a driving input and producing the occurrence of an event, and condition CM2 is not satisfied.  Scenario S3 indicates that some unique information is provided by $X_2$, along with a tiny amount of ShdM and Syn, and that increasing the strength of $X_1$ does increase the amount of transmitted information in S4. The line plot indicates that the ShdM and Syn components initially rise and then remain constant as $X_2$ increases towards 10. Thus there isn't the rise and fall in ShdM and Syn that one would expect to see with modulation of $X_2$ by $X_1$. This suggests that such modulation is not happening. 
\subsection*{Integration of the inputs, $X_1, X_2$?}
The surface plots of the mechanistic shared and synergy components in Fig. 10 suggest, for high
values of s1 and s2, when s1 and s2 are both increased, that the synergy and shared information
increase and start to level off. 
\begin{figure}[H]
\centering
\begin{tabular}{c}
\includegraphics[scale =0.5]{FIGURES/leg_mca.pdf} \\
 \includegraphics[scale=0.7]{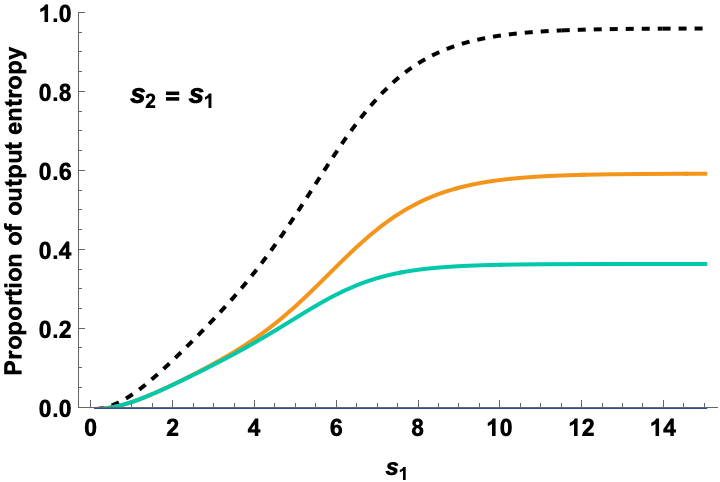}  
\end{tabular}
\caption{Plot of the normalised PID components along the line:  $s_2 = s_1$ using transfer function $T_6$ in $\eqref{T6}$. The dashed line traces the normalised joint mutual information. } \label{Fig13}
\end{figure}

The plot in Fig.~\ref{Fig13} show that this is the case along the line of $s1$-$s2$ equality. These plots suggest that signal integration is taking place. Thus, there is a suggestion of two
possibilities here:  (i) low $s_2$ with
unrestricted $s_1$ in which $X_2$ can modulate $X_1$, and (iii) high $s_1$ and $s_2$ where there is integration of
both inputs leading to increasing synergy and shared information coupled with decreasing or little
unique information of either kind.
\section*{A transfer function for divisive normalisation}
We take $X_2$ to be the integrated input from the outputs of neighbouring cells. A simple version of the transfer function \citet[Eq. (2)]{CH}  is given by
\begin{equation}
T(x_1, x_2) = a(x_1, x_2) = \frac{x_1^n}{k^n + x_1^n + x_2^n},  \label{DNgen}
\end{equation}
where $k, n$ are constants, which has a different form to the transfer function defined in (2) above. In order to accommodate this function and to use earlier results we make some modifications. First, replace the function $F$ in (2) by the identity mapping and replace the parameter $\mu$ by 0. Since $0 \leq a(x_1, x_2) <1$, the conditional output probabilities in Table 2 become simply $a_{00}, a_{10}, a_{02}, a_{12}$, respectively, and they represent in each case the event rate divided by the maximum event rate.  Then,  it follows that
\begin{equation}
a_{00} = 0, \,\,a_{10} = \frac{s_1^n}{k^n + s_1^n }, \,\,a_{02} =0, \,\,a_{12} =\frac{s_1^n}{k^n + s_1^n + s_2^n},
\end{equation}

The $\phi_i$ functions in (4)--(7) become
\begin{equation}
\phi_1(s_2) =0 ;\,\, \phi_2(s_1, s_2) = \mu a_{10} + \lambda a_{12};\,\, \phi_3(s_1) = \mu a_{10} ;\,\, \phi_4(s_1, s_2) = \lambda a_{12}.
\end{equation}

These $\phi_i$ functions must be used in (10)-(22) for the determination of information measures and PID components. Then, since $a_{12} > a_{00}$, we find from Theorem 1(a) that Unq$X_2$ = 0.

Taking $k=1, n=2$ in \eqref{DNgen}, we obtain 
\begin{equation}
T_7(x_1, x_2) = \frac{x_1^2}{1 + x_1^2 + x_2^2}.  \label{T7}
\end{equation}

\begin{figure}[H]
\centering
\includegraphics[scale =0.6]{FIGURES/leg_mca.pdf}

\begin{tabular}{cr}
\includegraphics[scale=0.5]{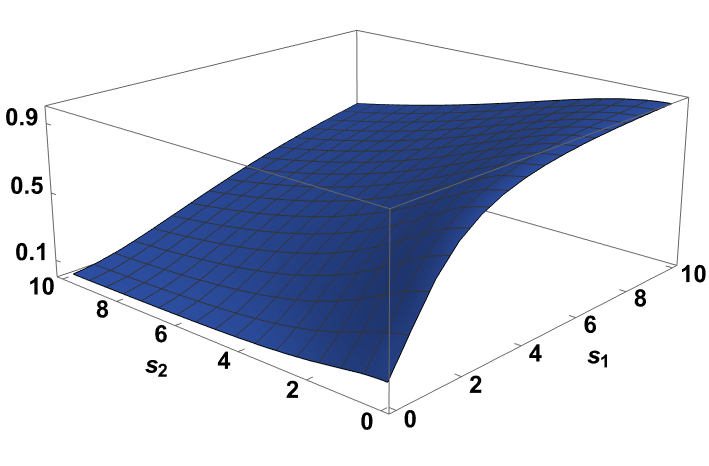} & \includegraphics[scale=0.5]{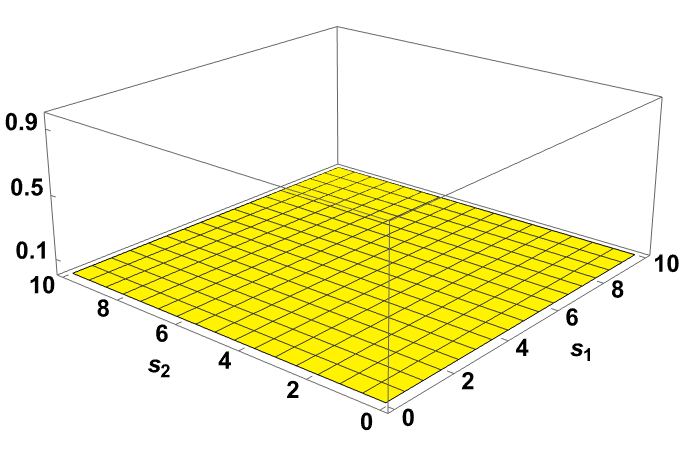}\\
\includegraphics[scale=0.5]{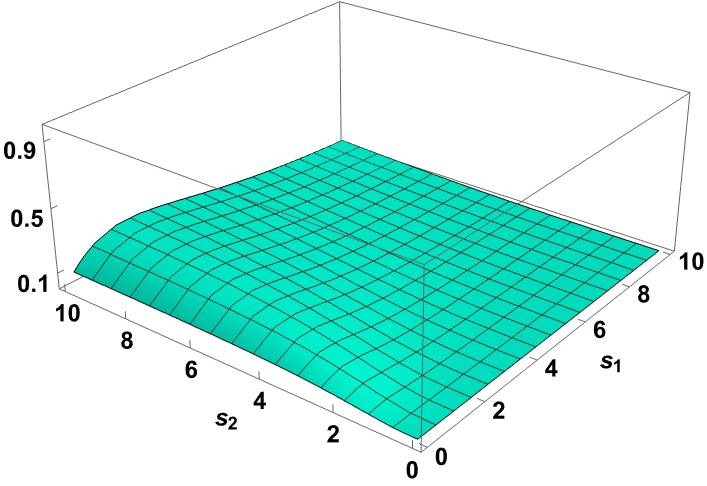} & \includegraphics[scale=0.5]{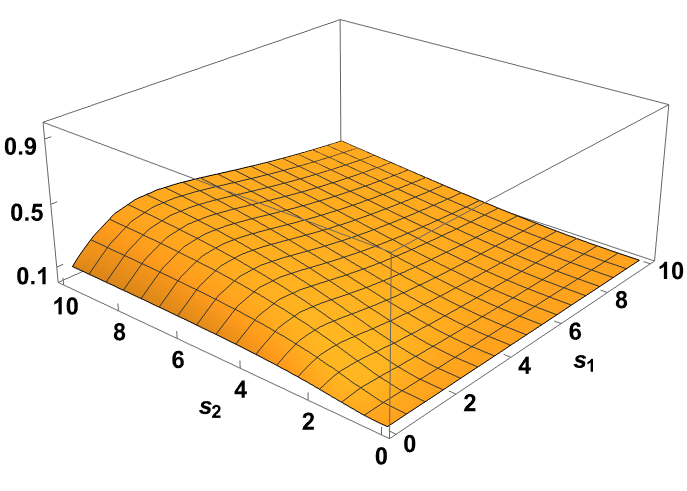} \\
\end{tabular}
\caption{Surfaces of the normalised PID components using the modulatory transfer function $T_7$, defined in $\eqref{T7}$} \label{DN1}
\end{figure}

Consider the surface plots in Fig.~\ref{DN1}. The plot of Unq$X_2$ illustrates the fact that this component is equal to zero for all
combinations of the input signal strengths, so $X_2$ holds no unique information about the output. As
the strength of input $X_1$ is increased from 0.1 to 10.0, more and more unique information held by
$X_1$ is transmitted; this component is the dominant one for large values of $s_1$. When the strength of
$X_1$ is low (0-4), we see that the synergy component rises and falls for most values of $X_2$. We also see that for large values
of $s_1$ and $s_2$ both the synergy and mechanistic shared components are close to zero; thus there is no
evidence of signal integration for transfer function $T_7$.

\begin{figure}[H]
\centering

\begin{tabular}{cc}
\includegraphics[scale =0.5]{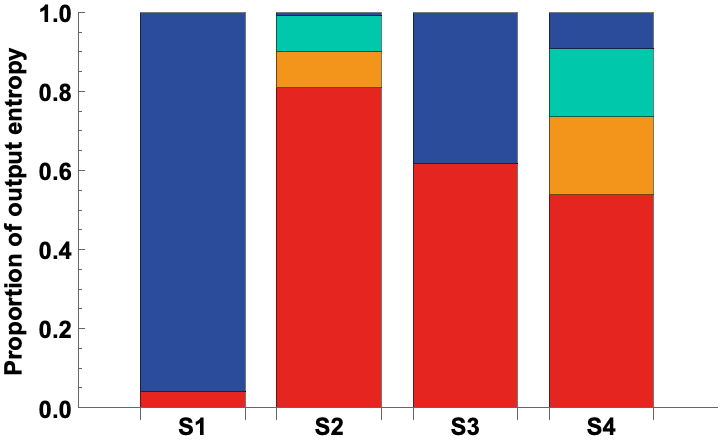}
& \includegraphics[scale=0.5]{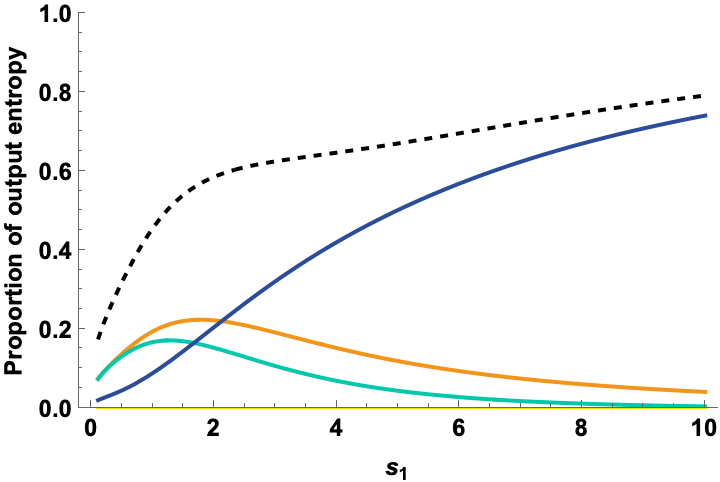}  \\
\end{tabular}
\caption{For transfer function $T_7$ in $\eqref{T7}$, normalised Ibroja PIDs for strength scenarios S\textsubscript{1} $(s_1 = 10, s_2 = 0.1)$, S\textsubscript{2} $(s_1 = 0.1, s_2=  10)$, S\textsubscript{3} $(s_1 = 1, s_2=  0.1)$ , S\textsubscript{4} $(s_1=  1, s_2 = 5)$;  The line plot shows the values of the normalised PID components as the $X_1$ input is increased from 0.1  to 10, and $X_2$,  fixed at 5. The dashed line traces the normalised joint mutual information. } \label{DN2}
\end{figure}

The barplot for Scenario $S_1$ in Fig.~\ref{DN2} indicates that almost all of the information is transmitted
as being unique information held by the $X_1$ input, when $X_1$ is strong and $X_2$ is close to zero. This
indicates that condition CM1 is satisfied. The barplot for scenario $S_2$ a little shared mechanistic and synergistic is transmitted, but no unique information provided by the context, with most of the available bandwidth left unused. 
Thus even when the context
$X_2$ is strong with almost zero drive from $X_1$ there is insufficient information transmitted for an event to
occur, supporting CM2. In Scenario S3, some unique information held by the input $X_1$ is transmitted
but none from input $X_2$, thus supporting condition CM3. The PID
for Scenario $S_4$ indicates that when the $X_1$ input is weak an increase in  $X_2$  results in
an increase in the transmission of information  in the form of synergistic and
mechanistic shared information as well as some unique information provided by input $X_1$. Thus condition
CM4 is satisfied. 
The line plot in Fig.~\ref{DN2} further illustrates the comments on the surfaces in Fig.~\ref{DN1}
 regarding the rise and fall of the information transmitted in the form of mechanistic shared and
synergistic information, suggesting that the condition for signal integration is not present. Hence,
we conclude that the transfer function $T_7$ is purely modulatory. 

\section*{Modulation and integration in a model layer 5b pyramidal cell}

\subsection*{The Data}
There are eleven  levels of basal amplitude, ranging from 0 (absent) to 1.0 nA, and eighteen levels of apical amplitude, ranging from 0 (absent) to 1.7 nA, giving 198 combinations of basal and apical input. For each combination, the model was run in 100 independent trials and a record made  of whether or not a burst was obtained. Thus by design, the basal input B and the apical input A both have a uniform distributions, and A and B are independent. Here, we use $B, A$ for $X_1, X_2$ and $b, a$ for $s_1, s_2$. See \citet{GKP}  for more detail.
\begin{figure}[H]
\centering
\includegraphics[scale =0.6]{FIGURES/leg_mca.pdf}

\begin{tabular}{cr}
\includegraphics[scale=0.4]{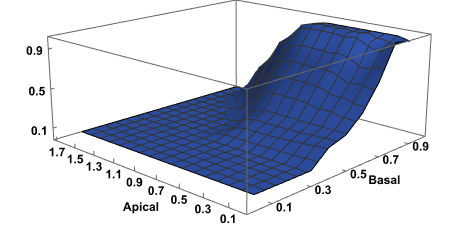} & \includegraphics[scale=0.4]{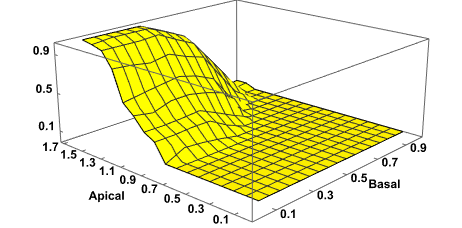}\\
\includegraphics[scale=0.4]{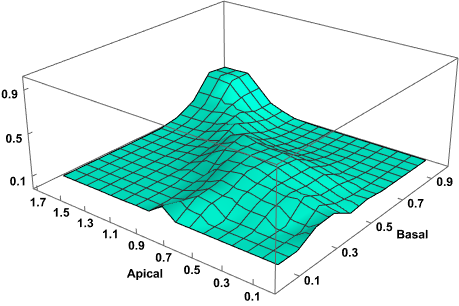} & \includegraphics[scale=0.4]{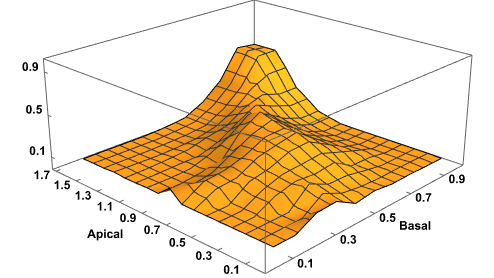} \\
\end{tabular}
\caption{Surfaces of the PID components for the HH data} \label{HHsurfs}
\end{figure}
\begin{figure}[H]
\centering
\begin{tabular}{cc}
\includegraphics[scale =0.45]{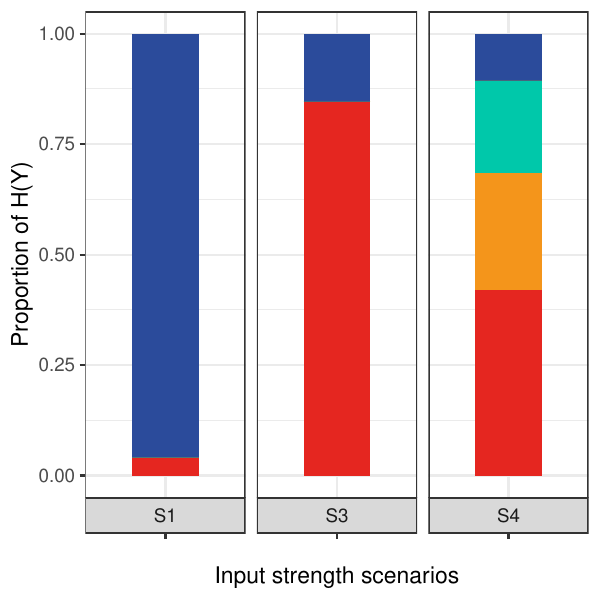}
 & \includegraphics[scale=0.45]{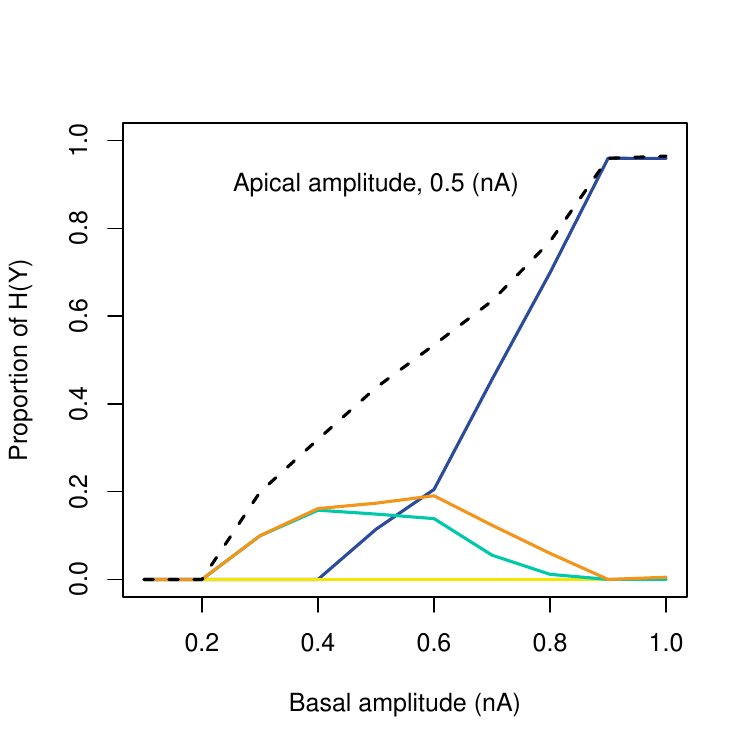}\\
\end{tabular}
\caption{Normalised Ibroja PIDs for strength scenarios S\textsubscript{1} (b = 0.9, a = 0.1), S\textsubscript{3} (b = 0.5, a =  0.1), S\textsubscript{4} (b = 0.5, a = 0.7). A line plot shows the values of the PID components as the basal amplitude is increased from 0.1 nA to 1.0 nA, and the  apical amplitude is fixed at  0.5 nA The dashed line traces the joint mutual information. } \label{AApids}
\end{figure}

\subsection*{Partial information decompositions}

Three-dimensional probability distributions were constructed, based on four of the experiments, and $A$ has levels 0 and a, $B$ has levels 0 and b, while $Y$ has levels 0 (no burst) or 1 (burst), ( b = 0.1,..., 1.0; a = 0.1, ..., 1.7) -- thus there are 170 such distributions. The Ibroja PID was applied to each distribution and computed using the {\em dit} package \citep{JEC}.  Since the inputs A and B are independent, the source-shared information component, ShdS, is equal to zero \citep{PPCP}, and so the shared information component comprises only mechanistic shared information, ShdM. PID components were normalised by dividing by the output entropy, $H(Y)$.

The PID surfaces over these basal and apical amplitude ranges are shown in Fig.16. They have the same general form as that given by a simple additive activation function (Fig. 5). This indicates there are regimes of contextual modulation and signal integration over restricted input amplitude ranges. We now discuss these regimes for this data.

\subsection*{Regime HL: Evidence for apical amplification}

Here the drive is the basal input, $B$, and the apical input, $A$, provides the context. The basal input, $B$ ranges from 0 to 1.0 nA, and $A$ ranges from 0 to 1.0 nA.  By the design of the experiments, it is not possible for the apical input to produce a burst on its own, while it is possible for the basal input to produce a burst on its own. The four signal-strength scenarios are defined as: S\textsubscript{1} (b = 0.9, a = 0.1), S\textsubscript{2} (b = 0, a = 1.0), S\textsubscript{3} (b = 0.5, a = 0.1), S\textsubscript{4} (b = 0.5, a = 0.7). Note that ``b = 0'' means that no basal current is applied.

The PID for Scenario S\textsubscript{1}  in Fig.~\ref{AApids} shows that almost all of the information is transmitted as being unique information provided by  the basal input, thus indicating that condition CM1 is satisfied: the basal information is sufficient to produce a burst, and that the apical input is not necessary to do so. Scenario S\textsubscript{2} cannot be checked in this regime by using PID. We know, by experimental design, that in regime HL it is not possible for the apical input to produce a burst by itself, and indeed when the basal input is absent and the apical input is equal to its highest possible value of 1.0 nA it is found that no burst is observed in each of the 100 trials. Thus condition CM2 is satisfied: the basal input is necessary to produce a burst and even the largest possible level of apical input is not sufficient. In Scenario S\textsubscript{3}, some unique information is provided by the basal input  but the apical input holds no unique information about bursting, thus supporting condition CM3. The PID for Scenario S\textsubscript{4} indicates that an increase in the apical amplitude, when the basal input is weak, results in an increase in the transmission of information about bursting in the form of synergistic, mechanistic shared information as well as unique information provided by the basal input.

The Line plot in Fig.~\ref{AApids} shows the profiles of the information components as the basal amplitude is increased from 0.1 nA to its maximum value at 1.0 nA. It is clear  that the transmission of synergistic and shared information rises and then falls as the basal amplitude is increased. We notice also that when the values of these PID components start to fall the transmission of unique information provided by the basal information rises and eventually is equal to the joint mutual information. This builds on what is seen in scenario S4, thus providing  evidence that $A$ can modulate information
transmission  for moderate values of $B$. On the contrary, transmission of unique information held by the apical amplitude is typically close to zero. The postulated conditions CM1--4 for contextual modulation in the form of apical amplification are thus satisfied.

\subsection*{Regime LH: Evidence for basal amplification}
Here the drive is the apical input, $A$, and the basal input, $B$, provides the context. The basal input, $B$ ranges from 0 to 0.5 nA, and $A$ ranges from 0 to 1.7 nA.  By the design of the experiments, it is not possible for the basal input to produce a burst on its own, while it is possible for the apical input to produce a burst on its own. The four signal-strength scenarios are defined as: S\textsubscript{1} (b = 0.1, a = 1.7), S\textsubscript{2} (b = 0.5, a = 0), S\textsubscript{3} (b = 0.2, a = 0.9), S\textsubscript{4} (b = 0.4, a = 0.9). Note that "a = 0" means that no apical  current is applied. 
\begin{figure}[H]
\centering
\includegraphics[scale =0.6]{FIGURES/leg_mca.pdf}
\begin{tabular}{cc}
\includegraphics[scale =0.45]{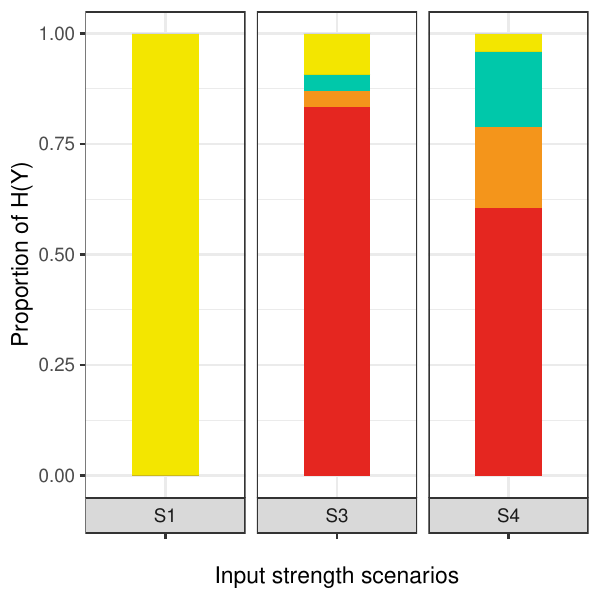}& 
 \includegraphics[scale=0.45]{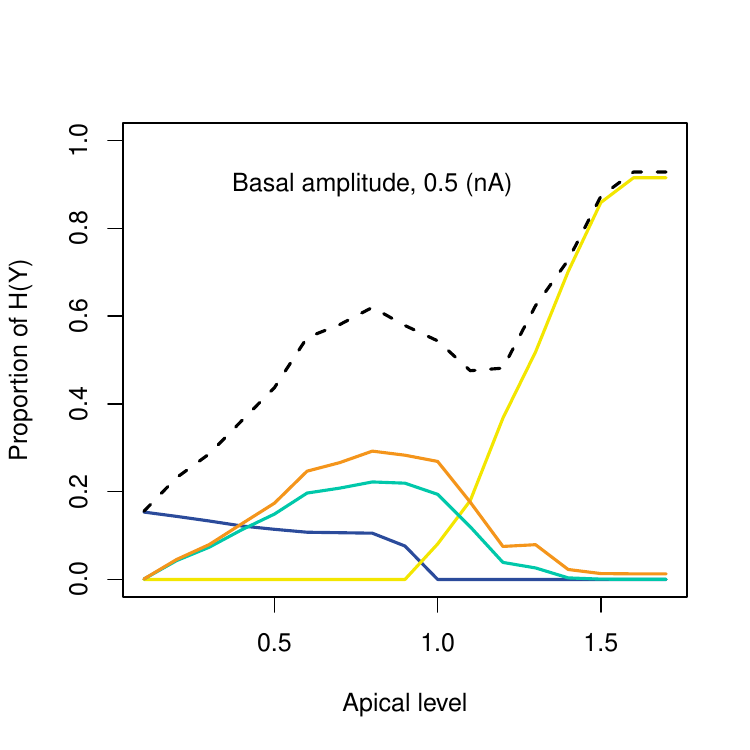} \\
\end{tabular}
\caption{Normalised Ibroja PIDs for strength scenarios S\textsubscript{1} (b = 0.1, a = 1.7), S\textsubscript{3} (b = 0.2, a =  0.9), S\textsubscript{4} (b = 0.4, a = 0.9). Line plots showing the values of the PID components as the apical amplitude is increased from 0.1 nA to 1.7 nA, with the basal amplitude fixed at  0.5 nA. The dashed line traces the joint mutual information. } \label{BApids}
\end{figure}

The PIDs in Fig.~\ref{BApids} are clear inverses of those in Fig.~\ref{AApids}. The PID for Scenario S\textsubscript{1}  in Fig. 3 shows that almost all of the information is transmitted as being unique information provided by  the apical input, thus indicating that condition CM1 is satisfied: the apical information is sufficient to produce a burst, and that the basal input is not necessary to do so. Scenario S\textsubscript{2} cannot be checked in this regime by using PID. We know, by experimental design, that in regime LH it is not possible for the basal input to produce a burst by itself, and indeed when the apical input is absent and the basal input is equal to its highest possible value of 0.5 nA it is found that no burst is observed in each of the 100 trials. Thus condition CM2 is satisfied: the apical input is necessary to produce a burst and even the largest possible level of basal input is not sufficient. In Scenario S\textsubscript{3}, some unique information provided by the apical input is transmitted, along with some synergy and mechanistic shared,  but the basal input holds no unique information about bursting, thus supporting condition CM3. The PID for Scenario S\textsubscript{4} indicates that an increase in the basal amplitude, when the apical input is weak, results in an increase in the transmission of information about bursting in the form of synergistic, mechanistic shared information as well as unique  information provided by  the apical input.

The Line plot in Fig.~\ref{BApids} shows the profiles of the information components as the apical amplitude is increased from 0.1 nA to its maximum value at 1.7 nA. It is clear in each case that the transmission of synergistic and shared information generally rises and then falls as the apical amplitude is increased. We notice also that when the values of these PID components starts to fall the transmission of unique  information provided by the apical input rises and eventually becomes equal to the joint mutual information. On the contrary, transmission of unique information provided by  the basal amplitude is typically close to zero, except for the case where the basal input is 0.5 nA when it decreases to zero. The postulated conditions CM1-4 for contextual modulation in the form of basal amplification are thus satisfied.
\subsection*{Regime HH: Evidence for basal--apical integration}

\begin{figure}[ht]
\centering
\includegraphics[scale =0.6]{FIGURES/leg_mca.pdf}

\begin{tabular}{cc}
\includegraphics[scale=0.45]{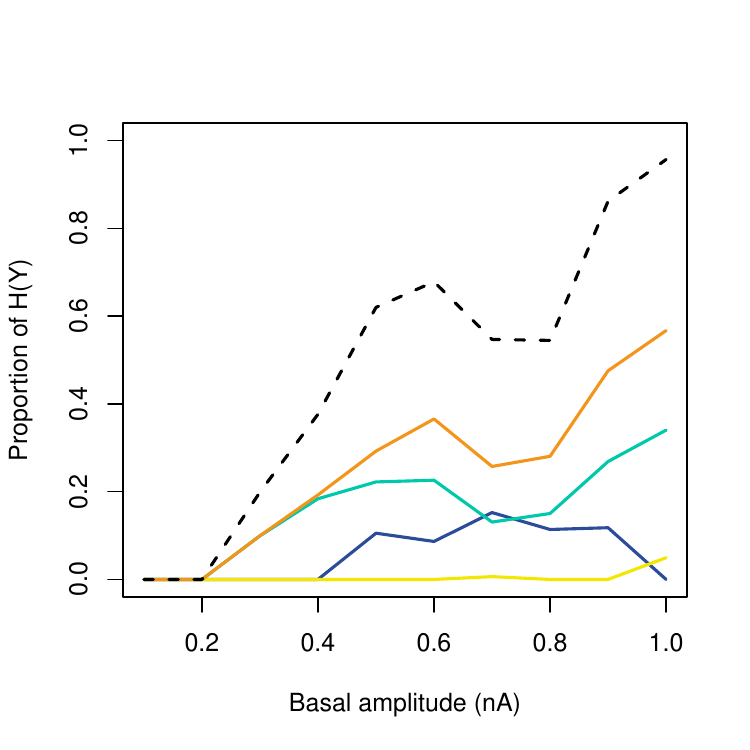} & \includegraphics[scale=0.45]{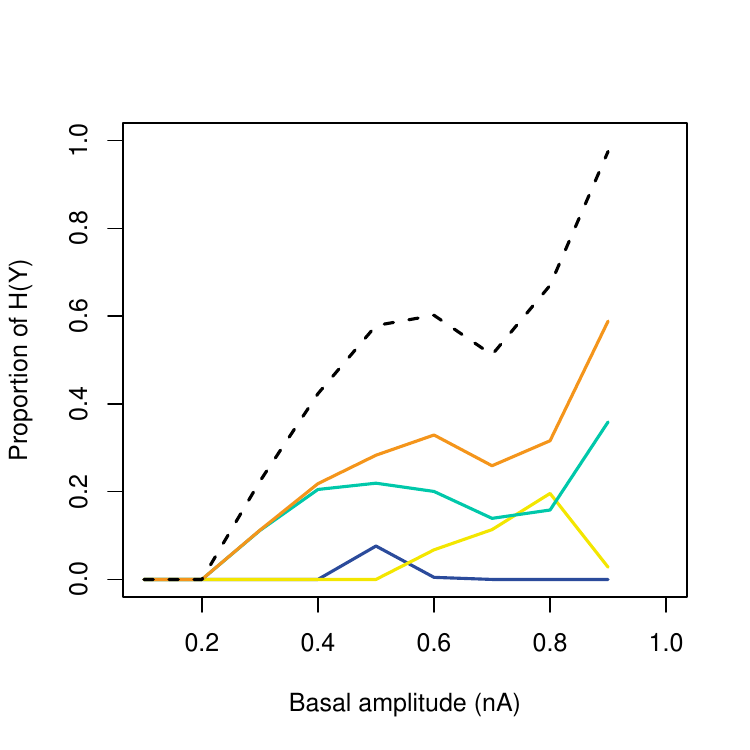}
\end{tabular}
\includegraphics[scale=0.45]{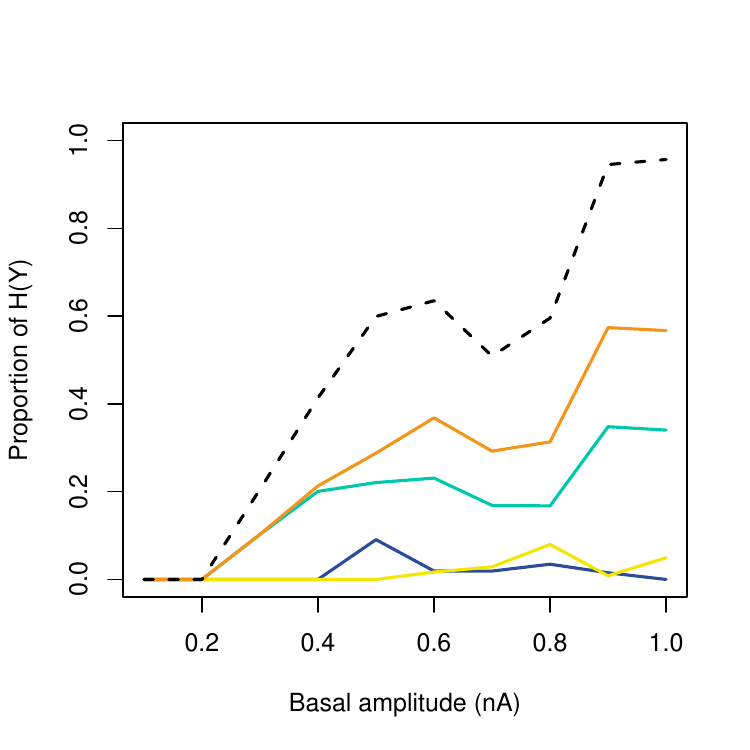}
\caption{Line plots of Ibroja PID components when the basal and apical amplitudes are increased simultaneously. The (b, a) amplitudes lie on the lines: $a = 1.6b$ (top-left), $a=1.8b$ (top-right) and $a = 1.7b$ (bottom). } \label{AAline}
\end{figure}

In the HH regime, both basal and apical amplitudes are allowed to reach high values (basal greater than 0.5  and apical greater than 1.0). As with the simple additive transfer function, as both basal and apical amplitudes are increased together, the shared and synergistic information components increase and eventually plateau, while the unique basal and apical components remain small (Fig. 16).

The line plots in Fig.~\ref{AAline} reveal that as both inputs are increased along lines of the form $a = k b$ the transmission of synergistic and shared information increases when both inputs are close to their maximum values. Correspondingly, the levels of transmitted unique information conveyed, respectively, by the apical input and the basal input decrease eventually or have low values. Thus the postulated condition for basal-apical integration is satisfied.

\section*{Conclusions}

The conditions for contextual modulation, together with four testing scenarios which make use of PID components, have been updated from their version in \citet{KP}.  They have been applied with several different transfer functions. A simple transfer function has been defined; the two inputs are fed into a general non-negative activation function, a threshold subtracted, and the result passed through a sigmoidal function to give the conditional probability of an event for given values of the two inputs. Exact results have been obtained for the information measures as well as the components of the Ibroja PID as continous functions of the signal strengths of the two inputs. Several general transfer functions have been defined based on additive, multiplicative, subtractive and divisive combinations of the two inputs, and deductions made regarding whether either or both of the unique informations are equal to zero. In the divisive case a particular form of a transfer function used for divisive normalisation \citep{CH} has been defined and exact results are available for it.

Several specific examples are discussed. Activation functions based on two different multiplicative  forms  are shown to provide purely modulatory transfer functions in which the driving input is modulated by the contextual input. In another illustration in which the activation function is a simple product of the two inputs, it is shown that the corresponding transfer function is purely integrative -- and not modulatory, as also found in \citet{KP}. 

The transfer function that is based on a simple addition of the two inputs provides interesting results. The Ibroja results reveal that different findings are obtained for the unique informations provided by drive and by context, depending on whether or not the driving signal strength is greater than  the contextual signal strength. Hence the restricted domains --  $s_1 > s_2$ and $s_1 < s_2$ -- are discussed separately. This approach was also motivated by the definition of different input regions in \citet{GKP}.  Evidence is provided that $X_1$ is modulated by $X_2$ when $s_1 > s_2$, whereas the reverse is true when $s_1 < s_2$. Also, when $s_1$ and $s_2$ are both high there is evidence of the integration of the two inputs. These conclusions are particularly rich especially when compared with the conclusions for this transfer function in the earlier paper \citep{KP}.  In this earlier work, the transfer function was considered only for the whole rectangular region for $(s_1, s_2)$ and the richness of these conclusions was missed. 

The same restricted $(s_1, s_2)$ domains were also considered when the transfer function was based on a simple subtraction of the two input signals. When the strength of the driving signal $s_1$ is greater than the strength of the inhibitory contextual signal $s_2$ the transfer function produces modulatory behaviour by depressing the probability of an event. When these relative signal strengths are reversed the conditions for contextual modulation are not all satisfied. As in the case of simple addition it is found that signal integration occurs when the signal strengths are both large.

The transfer function based on a version of divisive modulation revealed purely modulatory behaviour in that the integrated input from neighbouring cells reduced the probability of an event.

Finally, the information methodology was applied to the data from \citet{GKP}.  Evidence for apical amplification of the basal input in Regime HL, basal amplification of the apical input in regime LH, and basal-apical  integration in regime HH was presented. These conclusions are qualitatively similar to those for the activation function based on simple addition; the rectangular regimes HL and LH correspond to the triangular restricted domains $s_1 > s_2$ and $s_1 < s_2$, respectively, and the rectangular regime HH corresponds to the unrestricted rectangular domain $s_1 >0, s_2>0.$

\setcounter{table}{0}
\renewcommand{\thetable}{A\arabic{table}}

\renewcommand{\theequation}{A.\arabic{equation}}
\setcounter{equation}{0}
\section*{The Appendix}
\appendix
\section{Information measures }
For simplicity, we make the following  identifications for the expressions in $\eqref{p1}$--$\eqref{p4}$:
\begin{equation}
\phi_1 \equiv \phi_1(s_2), \quad \phi_2 \equiv \phi_2(s_1, s_2), \quad  \phi_3 \equiv \phi_3(s_1), \quad \phi_4 \equiv \phi_4(s_1, s_2). \label{A1}
\end{equation}
Recall from \eqref{phi} that
\begin{equation}
\phi_1 + \phi_2 = \phi_3 + \phi_4 \label{A1.5}
\end{equation}
We know by assumption that $s_1 >0, s_2>0$ and that $-1 < \rho < 1$. From Table 1, we deduce the following:
\begin{equation}
0 < \lambda < \tfrac{1}{2}, \quad 0 < \mu < \tfrac{1}{2}, \quad \lambda + \mu = \tfrac{1}{2}. \label{A2}
\end{equation} 
We define the function $h_2$ by
\begin{equation}
h_2(x) = - x \log x - (\tfrac{1}{2} -x) \log (1 - 2 x), \quad (0 < x < \tfrac{1}{2} 
\end{equation}
and show that $h_2(x) = \tfrac{1}{2} + \tfrac{1}{2} h(2 x)$, where the function $h$ is defined in $\eqref{h(x)}$, as follows.

\begin{align}
h_2(x) &= - x \log x - (\tfrac{1}{2} -x) \log (1 - 2 x) \nonumber \\
&=  -\frac{2 x}{2} \log \frac{2 x }{2} - \tfrac{1}{2} (1 - 2 x) \log \frac{(1-2x)}{2} \nonumber \\
&= x -\frac{2 x}{2} \log (2 x)  +\tfrac{1}{2} (1-2 x)- \tfrac{1}{2} (1 - 2 x) \log (1-2x) \nonumber\\
&= \tfrac{1}{2} + \tfrac{1}{2} h( 2 x) \label{A3}
\end{align}

We also state some properties of the strictly increasing, continuous cumulative probability distribution $F$:
\begin{equation}
0 < F(x) < 1, \quad x >y \iff F(x) > F(y), \quad x = y \iff F(x) = F(y), \label{A4}
\end{equation}
for real $x, y$, since $F$ is strictly increasing and also bounded by 0 and 1.
It then follows from  $\eqref{A1}$,  $\eqref{A2}$, $\eqref{A4}$, that 
\begin{equation}
 0 < \phi_k < \tfrac{1}{2}, \, (k =1,2,3,4). \label{A5}
 \end{equation}
 
 From Table 3, the marginal distributions of $(Y, X_1)$ and $(Y, X_2)$ can be obtained,  as follows.

\begin{table}[H]
\caption{Bivariate marginal distributions} \label{marg}
\begin{minipage}{0.4\textwidth}
\centering
\begin{tabular}{ccc} \toprule
$y$ & $x_1$ & $p(y, x_1)$ \\ \midrule
0 & 0 & $\tfrac{1}{2} - \phi_1$ \\
0 & $s_1$ &$ \tfrac{1}{2} -\phi_2$ \\
1 & 0 &  $\phi_1$ \\
1 & $s_1$ &$ \phi_2$ \\
\bottomrule
\end{tabular}
\end{minipage}
\begin{minipage}{0.4\textwidth}
\centering
\begin{tabular}{ccc} \toprule
$y$ & $x_2$ & $p(y, x_2)$ \\ \midrule
0 & 0 & $\tfrac{1}{2} - \phi_3$ \\
0 & $s_2$ &$ \tfrac{1}{2} -\phi_4$ \\
1 & 0 &  $\phi_3$ \\
1 & $s_2$ &$ \phi_4$ \\
\bottomrule
\end{tabular}
\end{minipage}
\end{table} 
From Table 3 or either of these tables, we see that the marginal distribution of $Y$ is: 
\begin{equation}
\Pr(Y=1) = \phi_1 + \phi_2, \quad  \Pr(Y=0) = 1 - \phi_1 - \phi_2. \label{A6}
\end{equation}
 Since $X_1$ and $X_2$ are uniform binary, and from $\eqref{A6}$, we have
 \begin{equation}
 H(X_1) =1, \quad H(X_2) =1, \quad H(Y) = h( \phi_1 + \phi_2). \label{A7}
 \end{equation}
From Table~\ref{tabRC}, we have that
\begin{equation}
H(X_1, X_2) = -2 \lambda \log \lambda - 2 \mu \log \mu. \label{A8}
\end{equation}
From Table~\ref{marg} we have
\begin{align*}
H(Y, X_1) &= - \phi_1 \log \phi_1 - (\tfrac{1}{2}- \phi_1) \log(\tfrac{1}{2} - \phi_1)  - \phi_2 \log \phi_2 - (\tfrac{1}{2} - \phi_2) \log(\tfrac{1}{2} - \phi_2)   \\
&= h_2(\phi_1) + h_2(\phi_2), \quad \text{by} \quad \eqref{A3} \\
&= 1 + \tfrac{1}{2} h(2 \phi_1) + \tfrac{1}{2} h( 2 \phi_2)
\end{align*}
Hence, using $\eqref{A7}$, it follows that
\begin{align}
         I(Y; X_1) &= H(Y) + H(X_1) - H(Y, X_1) \nonumber \\
         &= h(\phi_1 + \phi_2) - \tfrac{1}{2} h(2 \phi_1)    - \tfrac{1}{2} h(2 \phi_2).                     \label{A9}
\end{align}
By a similar argument,
\begin{equation}
I(Y, X_2) =  h(\phi_3 + \phi_4) - \tfrac{1}{2} h(2 \phi_3) - \tfrac{1}{2} h( 2 \phi_4). \label{A10}
\end{equation}
From $\eqref{A5}$, the expressions for $I(Y; X_1)$ and $I(Y; X_2)$ are well-defined.

From Table~\ref{tabYRC}, we have
\begin{align*}
H(Y, X_1, X_2) &= - \lambda F(a_{00}-\tau) \log \lambda F(a_{00}-\tau)    - \lambda ( 1 - F(-a_{00}\tau)) \log \lambda (1 - F(a_{00}-\tau))   \\
&\quad  - \mu F(a_{10}-\tau) \log \mu F(a_{10}-\tau)    - \mu ( 1 - F(a_{10}-\tau)) \log \mu (1 - F(a_{10} - \tau))   \\
&\quad  - \mu F(a_{02}-\tau) \log \mu F(a_{02}-\tau)    - \mu ( 1 - F(a_{02}-\tau)) \log \mu (1 - F(a_{02}-\tau))   \\
&\quad  - \lambda F(a_{12}-\tau) \log \lambda F(a_{12}-\tau)    - \lambda ( 1 - F(a_{12} -\tau)) \log \lambda (1 - F(a_{12}-\tau))  
\end{align*}

Consider the expression in the first row.
\begin{align*}
& - \lambda F(a_{00}-\tau) \log \lambda F(a_{00}-\tau)    - \lambda ( 1 - F(a_{00}-\tau)) \log \lambda (1 - F(a_{00}-\tau))   \nonumber \\
& = -\lambda F(a_{00}-\tau) \log \lambda  -\lambda F(a_{00}-\tau) \log F(-\tau)  - \lambda ( 1 - F(a_{00}-\tau)) \log \lambda  \nonumber \\ 
& \quad\quad\quad\quad  - \lambda ( 1 - F(a_{00}-\tau)) \log (1 - F(-a_{00}\tau)) \nonumber \\
& = -\lambda \log \lambda + \lambda h(F(a_{00}-\tau))
\end{align*}
Application of this argument to the expressions in the three other rows yields the following.
\begin{equation}
H(Y, X_1, X_2) = H(X_1, X_2) +  \lambda h(F(a_{00}-\tau)) + \mu h(F(a_{10}-\tau))+  \mu h(F(a_{02}-\tau)) + \lambda h(F(a_{12}-\tau))  \label{A12}
\end{equation}

Using $\eqref{A7}$ and $\eqref{A12}$, we find that
\begin{align}
I(Y; X_1, X_2) &= H(Y) + H(X_1, X_2) - H(Y, X_1, X_2)  \nonumber \\
&= h(\phi_1 + \phi_2) -  \lambda h(F(a_{00}-\tau)) - \mu h(F(ra_{10}-\tau))-  \mu h(F(a_{02}-\tau)) - \lambda h(F(a_{12}-\tau)).
\end{align}
Using the chain rule for mutual information  and an expression for mutual information \citep{CT} we have
\begin{align}
I(Y; X_1, X_2) &= I(Y; X_1) + I(Y; X_2|X_1) \label{A14}\\
&= I(Y; X_2) + I(Y;X_1|X_2) \label{A15}
\end{align}
and
\begin{equation}
H(Y) = I(Y; X_1, X_2) + H(Y|X_1, X_2) \equiv  I(Y; X_1, X_2) + H(Y)_\text{res}. \label{A16}
\end{equation}
Expressions for $I(Y;X_1|X_2), I(Y;X_2|X_1), H(Y)_\text{res}$ can be obtained from $\eqref{A14}$-$\eqref{A16}$.

\renewcommand{\theequation}{B.\arabic{equation}}
\setcounter{equation}{0}

\section{Proof of Theorem 1}

The defining equations of any partial information decomposition \citep{WB}  are
\begin{align}
I(Y; X_1) &= \text{Unq}X_1 + \text{Shd} \label{B1} \\
I(Y;X_2) &= \text{Unq}X_2 + \text{Shd} \label{B2}\\
I(Y; X_1, X_2) &= \text{Unq}X_1+ \text{Unq}X_2 + \text{Shd + Syn} \label{B3}
\end{align}
and two further equations can be deduced from them:
\begin{align}
I(Y;X_1|X_2) &= I(Y; X_1, X_2) - I(Y; X_2) = \text{Unq}X_1 + \text{Syn} \label{B4} \\
I(Y;X_2|X_1) &= I(Y;X_1, X_2) - I(Y; X_1) = \text{Unq}X_2 + \text{Syn}  \label{B5}
\end{align}
From Lemma 6 in \citet{Bert}, with altered notation,  a necessary and sufficient condition for Unq$X_2$ to vanish is that there exists a row stochastic matrix $S =[\sigma(x_1; x_2)]$ such that
\begin{equation}
p(y, x_2) = \sum_{\substack{x_1}} p(y, x_1) \,\sigma(x_1; x_2). \label{B6}
\end{equation}
From Table~\ref{marg}, and using $\eqref{B6}$, we have the following equations
\begin{align}
\tfrac{1}{2} - \phi_3 &= (\tfrac{1}{2} - \phi_1) \sigma(0; 0) + (\tfrac{1}{2} - \phi_2) \sigma(s_1; 0)  \label{B7}\\
\tfrac{1}{2}- \phi_4 &= (\tfrac{1}{2} - \phi_1) \sigma(0; s_2) +  (\tfrac{1}{2} - \phi_2)\sigma(s_1; s_2) \label{B8}\\
\phi_3 &=  \quad\quad\, \phi_1 \sigma(0; 0) + \phi_2 \sigma(s_1; 0)  \label{B9}\\
\phi_4 &= \quad\quad\, \phi_1 \sigma(0; s_2) + \phi_2 \sigma(s_1; s_2)  \label{B10}
\end{align}
Taking $\eqref{B7}$ and $\eqref{B9}$, we solve for $\sigma(0;0)$ and $\sigma(s_1;0)$.
\begin{align*}
\begin{bmatrix}
\sigma(0;0) \\
\sigma(s_1;0) \end{bmatrix}  &= \begin{bmatrix} \tfrac{1}{2} - \phi_1 & \tfrac{1}{2} - \phi_2 \\
\phi_1 & \phi_2 \end{bmatrix}^{-1} \begin{bmatrix} \tfrac{1}{2} - \phi_3 \\
\phi_3 \end{bmatrix}\\
&= \frac{1}{\Delta} \begin{bmatrix} \phi_2 & -(\tfrac{1}{2} - \phi_2) \\
-\phi_1 & (\tfrac{1}{2} - \phi_1) \end{bmatrix} \begin{bmatrix} \tfrac{1}{2} - \phi_3 \\
\phi_3 \end{bmatrix}\\
&= \frac{1}{\phi_2 -\phi_1} \begin{bmatrix} \phi_2 - \phi_3 \\ \phi_3 - \phi_1 \end{bmatrix}\\ 
&= \frac{1}{\phi_2 -\phi_1} \begin{bmatrix} \phi_4 - \phi_1 \\ \phi_2 - \phi_4 \end{bmatrix}
\end{align*}
since the determinant $\Delta$ is equal to $ \tfrac{1}{2} (\phi_2 - \phi_1)$, and using \eqref{A1.5}.
Repeating this argument using $\eqref{B8}$ and $\eqref{B10}$, we obtain
\begin{equation*}
\begin{bmatrix} 
\sigma(0;s_2) \\
\sigma(s_1;s_2) \end{bmatrix}  =   \frac{1}{\phi_2 -\phi_1} \begin{bmatrix} \phi_2 - \phi_4 \\ \phi_4 - \phi_1 \end{bmatrix}
\end{equation*}
We now require to discuss whether
\begin{equation*}
S = \begin{bmatrix} \sigma(0;0) & \sigma(0;s_2) \\
\sigma(s_1;0) & \sigma(s_1;s_2) \end{bmatrix} = \frac{1}{\phi_2 - \phi_1} \begin{bmatrix} \phi_4 - \phi_1 & \phi_2 - \phi_4 \\
\phi_2 -\phi_4 & \phi_4 - \phi_1   \end{bmatrix}
\end{equation*}
is a row-stochastic matrix. Using $\eqref{p1}$- $\eqref{p4}$ and $\eqref{A1}$,  we check the signs of the various expressions.
\begin{align*}
\phi_2 - \phi_1 &= \left( \lambda ( F(a_{12} -\tau) - F(a_{00}-\tau)) \right) + \left( \mu (F(a_{10} -\tau) - F(a_{02} - \tau)) \right)\\
\phi_4 - \phi_1 &= \lambda (F(a_{12} -\tau) - F(a_{00}-\tau))\\
\phi_2 - \phi_4 &= \mu (F(a_{10}  -\tau) - F(a_{02}-\tau))
\end{align*}
By assumption, $a_{12} > a_{00}$ and $F$ is strictly increasing so  the first term in $\phi_2 - \phi_1$ is positive. Therefore, using $\eqref{A4}$,  $\phi_2 - \phi_1$ is positive when $a_{10}\geq a_{02}$. Also $\phi_4 - \phi_1 >0$, and $\phi_2 - \phi_4$ is non-negative when $a_{10} \geq a_{02}$. Therefore the entries of $S$ are non-negative and clearly the row sums of $S$ are equal to 1. 
\newline

\noindent Hence, when $a_{10} >a_{02}$ and also when $a_{10} =a_{02}$ it follows that $S$ is row-stochastic. Hence, when $a_{10} > a_{02}$ and also when $a_{10}=a_{02}$, it follows that Unq$X_2$ =0. Using (B.1)-(B.5)  we deduce that the exact Ibroja PID when $a_{10} > a_{02}$ is:
\newline

Unq$X_1$ = $I(Y;X_1) - I(Y;X_2)$, Unq$X_2$ =0, Shd = $I(Y;X_2)$, Syn = $I(Y;X_2|X_1),$
which are written as functions of $(s_1, s_2)$ in $\eqref{iyr}$, $\eqref{iyc}$ and $\eqref{iy1G2}$.

This argument can be repeated  in the case where $ a_{10} \leq a_{02}$ by swapping the roles of $X_1$ and $X_2$ in $\eqref{B6}$ -- $\eqref{B10}$. The details are very similar to the above and provide the following result.   Hence, when $a_{10} < a_{02}$ and when $a_{10}=a_{02}$, it follows that Unq$X_1$ =0. The exact Ibroja PID when $a_{10} < a_{02}$ is:
\newline

Unq$X_1$ =0;  Unq$X_2$ =  $I(Y;X_2) - I(Y;X_1)$, Shd = $I(Y;X_1)$, Syn = $I(Y;X_1|X_2),$
which are written as functions of $(s_1, s_2)$ in $\eqref{iyr}$, $\eqref{iyc}$, $\eqref{iy1G2}$ and $\eqref{iy2G1}$.

Combining  the results for $a_{10} \geq a_{02}$ and $a_{10} \leq a_{02}$, and using $\eqref{B1}$--$\eqref{B5}$, the exact Ibroja PID for the case when $a_{10} = a_{02}$ is
\newline

Unq$X_1$ =0;  Unq$X_2$ =  0, Shd = $I(Y;X_1)$, Syn = $I(Y;X_1|X_2)$, which are written as functions of $(s_1, s_2)$ in $\eqref{iyr}$ and $\eqref{iy1G2}$. Note that in this case
\begin{equation*}
I(Y; X_1) = I(Y;X_2) \quad \text{and} \quad I(Y;X_1|X_2) = I(Y;X_2|X_1).
\end{equation*}

The proof of part (b) of Theorem 1 is similar to that given above by swapping the roles of $X_1$ and $X_2$ in (B.6).
\renewcommand{\theequation}{C.\arabic{equation}}
\setcounter{equation}{0}

\section{Proof of Theorem 2}
Suppose first that $\rho = +1$. Then from $\eqref{A2}$ we have that $ \lambda = \tfrac{1}{2}, \mu =0$. Hence from $\eqref{p1}$--$\eqref{p4}$  it follows that 
\begin{equation}
\phi_1 = \tfrac{1}{2} F(a_{00}-\tau), \quad \phi_2 = \tfrac{1}{2} F(a_{12}-\tau),\quad  \phi_3 = \tfrac{1}{2} F(a_{00}-\tau), \quad \phi_4 = \tfrac{1}{2} F(a_{12}-\tau).
\end{equation}
Using (C.1), and (13)--(15), we find that $I(Y;X_1), I(Y;X_2), I(Y;X_1, X_2)$ are each equal to
\begin{equation}
h( \tfrac{1}{2} ( F(a_{00}-\tau) + F(a_{12}-\tau)) - \tfrac{1}{2} h(F(-a_{00}\tau))    - \tfrac{1}{2} h(F(a_{12}-\tau)).
\end{equation}
Hence, from $\eqref{B4}$, $\eqref{B5}$,  $I(Y;X_1|X_2)$ and $I(Y;X_2|X_1)$ are equal to zero, and  it follows that Unq$X_1$ = Unq$X_2$ = Syn = 0, and Shd is equal to the expression in (C.2).

In the case of $\rho = -1$, a similar argument leads to the following expression for Shd.
\begin{equation*}
    h( \tfrac{1}{2} ( F(a_{10}-\tau) + F( a_{02}-\tau)) - \tfrac{1}{2} h(F(a_{10}-\tau))    - \tfrac{1}{2} h(F(a_{02} -\tau))
    \end{equation*}
    In both cases,  it follows that Shd = ShdS because the non-negative ShdM component is bounded above by the synergy term (Pica et al., 2017) which is equal to zero.


\begin{thebibliography}{99.}

\bibitem[Bertschinger et al.(2014)]{Bert} Bertschinger, N.; Rauh, J.; Olbrich, E.; Jost, J.;  Ay, N. Quantifying Unique Information. {\em Entropy} {\bf 2014}, {\em 16}, 2161--2183.

\bibitem[Brozovi{\'c} et al.(2008)]{BAA} Brozovi{\'c}, M.; Abbott, L.F.; Andersen, R.A. Mechanism of gain modulation at single neuron and network levels. {\em J Comput Neurosci} {\bf 2008}, {\em 25}, 158--168.

\bibitem[Carandini and Heeger(2012)]{CH} Carandini, M.; Heeger, D.J. Normalization as a canonical computation. {\em Nat. Rev. Neurosci.} {\bf 2012}, {\em 13}, 51--62.

\bibitem[Cover and Thomas(1991)]{CT} Cover, T.M.;  Thomas, J.A.  Elements of Information Theory; Wiley-Interscience: New York, NY, USA, 1991.

\bibitem[Graham et al.(2025)]{GKP} Graham, B.P.; Kay, J.W.; Phillips, W.A. Context-Sensitive Processing in a Model Neocortical Pyramidal Cell With Two Sites of Input Integration. {\em Neural Computation} {\bf 2025}, {\em 37(4)}, 588--634.

\bibitem[Griffith and Koch(2014)]{GK} Griffith, V.;  Koch, C.; Griffith, V. Quantifying synergistic mutual information. In {\em Guided Self-Organization: Inception. Emergence, Complexity and Computation}; Springer: Berlin/Heidelberg, Germany, 2014; Volume 9, pp.~159--190.

\bibitem[Harder et al.(2013)]{HSP} Harder, M.; Salge, C.; Polani, D. Bivariate measure of redundant information. {\em Phys. Rev. E} {\bf 2013}, {\em 87}, doi:10.1103/PhysRevE.87.012130.

\bibitem[James et al.(2018)]{JEC} James, R. G.; Ellison C. J.; Crutchfield J.P. a Python package for discrete information theory. {\em The Journal of Open Source Software} {\bf 2018}, 25:738. https://doi.org/10.21105/joss.00738. 

\bibitem[Kay et al.(2017)]{KIDP} Kay, J. W.;  Ince, R. A. A.;  Dering, B.;  Phillips, W. A. Partial and Entropic Information Decompositions of a Neuronal Modulatory Interaction. {\em Entropy} {\bf 2017}. 19(11), 560.  doi:10.3390/e19110560.


\bibitem[Kay and Phillips(2020)]{KP} Kay, J. W.;  Phillips, W. A. Contextual Modulation in Mammalian Cortex is Asymmetric. {\em Symmetry} {\bf 2020}, 12, 815.

\bibitem[Larkum(2013)]{Lark} Larkum, M. A cellular mechanism for cortical associations: An organizing principle for the cerebral cortex. {\em Trends Neurosci.} {\bf 2013}, {\em 36}, 141--151.

\bibitem[Phillips(2017)]{WAP} Phillips, W.A. Cognitive functions of intracellular mechanisms for contextual amplification. {\em Brain and Cognition} {\bf 2017}, {\em 112}, 39--53.

\bibitem[Phillips et al.(2015)]{PCS} Phillips, W.A.; Clark, A.; Silverstein, S.M. On the functions, mechanisms, and malfunctions of intracortical contextual modulation. {\em Neurosci. Biobehav. Rev.} {\bf 2015}, {\em 52}, 1--20.

    
    \bibitem[Phillips et al.(1995)]{PKS} Phillips, W.A.; Kay, J.;  Smyth, D. The discovery of structure by multi-stream networks of local processors with contextual guidance. {\em Netw. Comput. Neural Syst.} {\bf 1995}, {\em 6}, 225--246.



\bibitem[Pica et al.(2017)]{PPCP} Pica, G.; Piasini, E.; Chicharro, D, ; Panzeri, S. Invariant components of synergy, redundancy, and unique information. {\em Entropy} {\bf 2017}, {\em 19}, 451, doi:10.3390/e19090451.


\bibitem[Silver(2010)]{RAS} Silver, R.A. Neuronal arithmetic. {\em Nat. Rev. Neurosci.}  {\bf 2010}, {\em 11}, 474--489. 

\bibitem[Wibral et al.(2017)]{WPKLP} Wibral, M.; Priesemann, V.; Kay, J.W.; Lizier, J.T.; Phillips, W.A. Partial information decomposition as a unified approach to the specification of neural goal functions. {\em Brain  Cognit.} {\bf 2017}, {\em 112}, 25--38.

\bibitem[Williams and Beer(2010)]{WB}Williams, P.L.;  Beer, R.D. Nonnegative Decomposition of Multivariate Information. \emph{arXiv}  \textbf{2010}, arXiv:1004.2515.  

\end{thebibliography}
     \end{document}